\documentclass{article}

\usepackage[english]{babel}

\usepackage[letterpaper,top=2cm,bottom=2cm,left=3cm,right=3cm,marginparwidth=1.75cm]{geometry}

\usepackage{amsmath, amsfonts}
\usepackage{graphicx}
\usepackage{siunitx}
\usepackage{xcolor}
\usepackage{listings}
\usepackage{mathtools}

\usepackage{doi}

\usepackage{authblk}

\usepackage{hyperref}

\definecolor{codegreen}{rgb}{0,0.6,0}
\definecolor{codegray}{rgb}{0.5,0.5,0.5}
\definecolor{codepurple}{rgb}{0.58,0,0.82}
\definecolor{backcolour}{rgb}{0.95,0.95,0.92}
\lstdefinestyle{pystyle}{
    backgroundcolor=\color{backcolour},   
    commentstyle=\color{codegreen},
    keywordstyle=\color{magenta},
    numberstyle=\tiny\color{codegray},
    stringstyle=\color{codepurple},
    basicstyle=\ttfamily\footnotesize,
    breakatwhitespace=false,         
    breaklines=true,                 
    captionpos=b,                    
    keepspaces=true,                 
    numbersep=5pt,                  
    showspaces=false,                
    showstringspaces=false,
    showtabs=false,                  
    tabsize=2,
    language=python
}
\lstdefinestyle{defaultstyle}{
    backgroundcolor=\color{backcolour},   
    commentstyle=\color{codegreen},
    keywordstyle=\color{magenta},
    numberstyle=\tiny\color{codegray},
    stringstyle=\color{codepurple},
    basicstyle=\ttfamily\footnotesize,
    breakatwhitespace=false,         
    breaklines=true,                 
    captionpos=b,                    
    keepspaces=true,                 
    numbersep=5pt,                  
    showspaces=false,                
    showstringspaces=false,
    showtabs=false,                  
    tabsize=2,
}

\newcommand{\half}{\frac{1}{2}}

\newcommand{\m}{\mathcal}

\newcommand{\la}{\langle}
\newcommand{\ra}{\rangle}

\newcommand{\st}{$_2$}

\newcommand{\maxkcut}{Max-$k$-Cut}

\newcommand{\intellabscali}{Intel Labs, Santa Clara, CA 95054, USA}
\newcommand{\intellabsoreg}{Intel Labs, Hillsboro, OR 97124, USA}
\newcommand{\inteloreg}{Intel Corporation, Hillsboro, OR 97124, USA}
\newcommand{\oxfordmatsci}{Department of Materials, University of Oxford, Oxford OX1 3PH, UK}
\newcommand{\oxfordpchem}{Physical \& Theoretical Chemistry Laboratory, University of Oxford, Oxford, OX1 3QZ, UK}
\newcommand{\texasam}{Department of Chemistry, Texas A\&M University, College Station, TX 77843, USA}
\newcommand{\nasaames}{NASA Ames Research Center, Moffett Field, CA 94035, USA}
\newcommand{\usra}{Universities Space Research Association, Mountain View, CA 94035, USA}
\newcommand{\purdue}{Davidson School of Chemical Engineering, Purdue University, West Lafayette, IN 47907, USA}
\newcommand{\lbnl}{Lawrence Berkeley National Lab, Berkeley, California 94720}
\newcommand{\nersc}{National Energy Research Scientific Computing Center (NERSC), Lawrence Berkeley National Laboratory, Berkeley, CA 94720, USA}
\newcommand{\sandia}{Sandia National Laboratories, Albuquerque, NM 87185, USA}
\newcommand{\azulenelabs}{Azulene Labs, San Francisco, CA 94115, USA}

\title{HamLib: A library of Hamiltonians for benchmarking quantum algorithms and hardware}

\date{}

\author[1,2]{Nicolas PD Sawaya\thanks{nicolas@azulenelabs.com}}
\affil[1]{\azulenelabs}
\affil[2]{\intellabscali}

\author[3]{Daniel Marti-Dafcik}
\affil[3]{\oxfordpchem}

\author[4]{Yang Ho}
\affil[4]{\sandia}

\author[5]{Daniel P Tabor}
\affil[5]{\texasam}

\author[6,7,8]{David E Bernal Neira}
\affil[6]{\nasaames}
\affil[7]{\usra}
\affil[8]{\purdue}

\author[4]{Alicia B Magann}

\author[9]{Shavindra Premaratne}
\affil[9]{\intellabsoreg}

\author[2]{Pradeep Dubey}

\author[9]{Anne Matsuura}

\author[10]{Nathan Bishop}
\affil[10]{\inteloreg}

\author[11]{Wibe A de Jong}
\affil[11]{\lbnl}

\author[12]{Simon Benjamin}
\affil[12]{\oxfordmatsci}

\author[4]{Ojas Parekh}

\author[6]{Norm M Tubman}

\author[13]{Katherine Klymko\thanks{kklymko@lbl.gov}}
\affil[13]{\nersc}

\author[13]{Daan Camps\thanks{dcamps@lbl.gov}}

\begin{document}

\maketitle

\begin{abstract}
In order to characterize and benchmark computational hardware, software, and algorithms, it is essential to have many problem instances on-hand. This is no less true for quantum computation, where a large collection of real-world problem instances would allow for benchmarking studies that in turn help to improve both algorithms and hardware designs. To this end, here we present a large dataset of qubit-based quantum Hamiltonians. The dataset, called HamLib (for Hamiltonian Library), is freely available online and contains problem sizes ranging from 2 to 1000 qubits. HamLib includes problem instances of the Heisenberg model, Fermi-Hubbard model, Bose-Hubbard model, molecular electronic structure, molecular vibrational structure, MaxCut, Max-$k$-SAT, Max-$k$-Cut, QMaxCut, and the traveling salesperson problem. The goals of this effort are (a) to save researchers time by eliminating the need to prepare problem instances and map them to qubit representations, (b) to allow for more thorough tests of new algorithms and hardware, and (c) to allow for reproducibility and standardization across research studies.
\end{abstract}

\tableofcontents

\section{Scope and Preliminaries}

\subsection{Motivation}

Large datasets of problem instances have long been useful for the analysis of computer hardware, software, and algorithms. 
For instance, ImageNet \cite{deng2009imagenet} is a massive repository of images that has facilitated the testing of many deep learning packages. Another example is computational chemistry and materials science, where extensive datasets (e.g. the Protein Data Bank \cite{berman2000protein}, the Materials Project \cite{jain2013matproj} and QM9 \cite{ramakrishnan2014qm9}) improve ease of testing new algorithms and software approaches, while providing standardization across the field. Well-defined datasets or problem instances may in turn be used to define benchmarking suites such as MLPerf \cite{mattson2020mlperf} and LINPACK \cite{dongarra2003linpack,mohammadi2018linpack}.

Though there has been progress in introducing benchmarks in the quantum computing community \cite{parekh2016benchmarking,chen2022veriqbench,li2022qasmbench,Lubinski21_bench,cornelissen2021scalable,tomesh2022supermarq}, there is not yet a topically broad database of problem instances. Having such a dataset would be convenient for many reasons. For instance, when researchers wish to test a novel Hamiltonian simulation algorithm \cite{lloyd1996universal,whitfield2011simulation,low2017optimal,childs2021theory} for chemistry, they must first go through the tedious and non-trivial process of preparing a set of chemical Hamiltonians on their own. It would be useful for the researcher to have these preparatory steps done ahead of time, so that they may spend more of their efforts on algorithm or hardware design.

There are three primary motivations behind creating a dataset of Hamiltonians with broad coverage in application area and in problem difficulty. First, such a library can save substantial labor time. For example, a researcher wanting to test their new quantum chemistry algorithms will not have to learn the minutia of electronic structure, install and run various packages, choose a representative test set, and debug inevitable software difficulties.
In turn, this allows for resources and time being devoted to to the more interesting aspects of algorithm and software development.

Second, a large Hamiltonian library allows for more thorough testing. For example, if one is performing numerical tests on a new Hamiltonian simulation algorithm, immediately being able to run it on a very broad class of problems (as opposed to only e.g. a toy spin model and two to three molecules) allows for a stronger understanding of when the algorithm performs well.

Third, such a library allows for reproducibility and standardization across research studies. It is more straightforward to make fair comparisons between two algorithms if they are benchmarked using the exact same problem sets.

In order to facilitate practical benchmarking of quantum algorithms, software, and hardware, in this work we introduce an extensive dataset of many quantum \textit{problem Hamiltonians} from a variety of fields related to condensed matter physics, chemistry, and classical optimization. We call the dataset HamLib (for Hamiltonian Library) version 1. We focus primarily on presenting a wide range of Hamiltonians that may be used in various contexts; the definition of proper benchmarks---which generally require defining both a problem and an algorithm \textit{in addition to} a dataset---will not be our primary goal. However, we include broad discussion of benchmarking in Section \ref{sec:bench}.

\begin{figure}
    \centering
    \includegraphics[width=.7\textwidth]{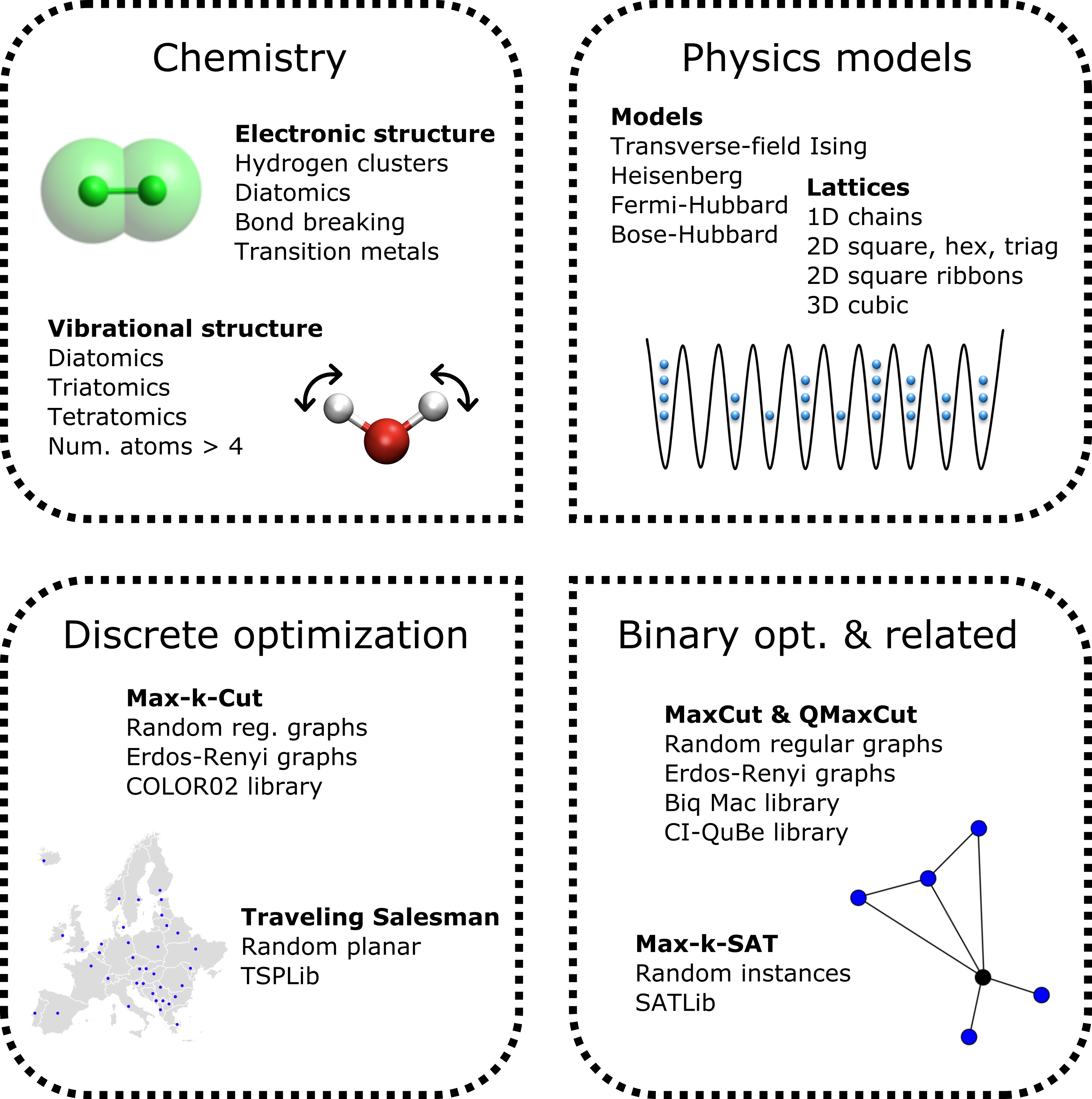}
    \caption{The four categories of qubit Hamiltonians included in HamLib. In the area of chemistry we include real-world accurate Hamiltonians for both electronic and vibrational structure. The condensed matter dataset includes four commonly studied models defined on a variety of lattice topologies, with and without periodic boundary conditions. We include Max-$k$-SAT and MaxCut in the binary combintorial problems, as well as QMaxCut, a quantum analogue with interesting mathematical properties. For discrete combinatorics we include both Max-$k$-Cut and the traveling salesperson problem. 
    }
    \label{fig:problems}
\end{figure}

\subsection{Attributes of this library}\label{sec:attributes}

There were several principles that guided the curation of this library. Although choosing the contents for any dataset is somewhat arbitrary, we believe that HamLib's contents are well-motivated, while showing exceptional breadth in terms of problem size, complexity, mathematical properties, and field of study.

We strive to create datasets that were ``well-spaced" in terms of qubit counts. This is important for multiple reasons. First, strong simulation of quantum algorithms cannot be performed past 40 to 50 qubits; hence in order to properly study scaling, one should be studying problem instances for every few qubits. Even for cases where tensor networks are used to simulate 100s of qubits, or where quantum compilers are benchmarked on even more qubits, problem sizes in the context of early quantum computing are still much smaller than typical classical datasets. For instance, the classical benchmark Graph 500 \cite{murphy2010graph500} contains graph problem instances of more than $10^{40}$ vertices. Second, because current quantum hardware is still limited in size, small spacings are necessary in order to test as much of the device as possible. For instance, if an experimental quantum computer has exactly 24 qubits, we want to ensure that several problem instances from HamLib can be used to test the full device.

A main focus of HamLib is the curation of \textit{real-world} problems, with parameters and properties as similar as possible to those encountered in some real scientific or industrial applications. We include high-quality and meticulously prepared problems in molecular electronic structure and vibrational structure, while also curating real-world datasets for combinatorial problems such as routing, by using distance matrices from real cities \cite{reinelt1991tsplib}. It is our suspicion %
that the reason many quantum algorithms are often tested only on simpler condensed matter models or on random graphs is simply that it is time-consuming to prepare problems that are more representative of the real world. However, lattice models and random instances are equally vital for benchmarking and scientific understanding, and as such much of our dataset consists of toy problems as well.

Another essential facet of HamLib is that all problem instances have already been mapped to qubits, i.e. they have already been mapped to a Pauli representation of the form
\begin{equation}
H_{\text{encoded}} = \sum_i c_i \bigotimes_k \{\sigma_{ik}\}%
\end{equation}
where $\sigma_{ik}$ is a one-qubit Pauli or identity operator, i.e. $\sigma_{ik} \in \{I,X,Y,Z\}$, and $c_i$ is a real scalar. It is our hope that this eliminates substantial labor for some researchers. The user can simply download the qubit representations and immediately use them in a qubit-based computer or simulator. However, we caution against a black box approach. It will often be important to understand the difference between (for example) the various fermion-to-qubit mappings or the various integer-to-qubit mappings, before interpreting the results of a particular benchmark. The original ``unencoded'' Hamiltonians are included where appropriate, so that the user can implement alternative encoding approaches not considered here. Further, some auxiliary information is included, such as approximate ground states or operators that aid in implementing certain quantum algorithms.

We attempt to achieve broad coverage in terms of the mathematical properties of the Hamiltonians. A broad range of localities (i.e. Pauli weights) are present, as the many mapping choices for fermionic, vibrational, bosonic, and combinatorial problems lead to localities of anywhere from 2 to $N$, where $N$ is the number of qubits. Qubit connectivites vary widely as well, as we consider many graph types for combinatorial problems, and several grid dimensionalities and types in condensed matter. Finally, we note that a broad range of Hamiltonian norms are present---this is important because the formal complexity of many algorithms (especially in Hamiltonian simulation \cite{childs2021theory,low2022trotter}) depend explicitly on such norms. %

A main use case we have in mind for this dataset is comparing quantum approaches against each other, not just comparing quantum algorithms/computers to classical ones. Hence there are entire problem classes for which the task is either certain or likely to scale polynomially on a classical computer (e.g. ground state finding for the 1D transverse-field Ising model)---such problem instances are still useful for quantum-to-quantum comparisons for quantum algorithms, quantum compilers, and quantum computers. However, for many portions of HamLib, it would certainly be a reasonable use of the dataset to compare scaling behavior in quantum versus classical methods, and we encourage this as use case as well.

Some comments are merited on why certain problem areas are \textit{not} included in HamLib. We have focused exclusively on problems that are commonly represented \textit{as Hamiltonians}, largely for the purpose of saving time for researchers. For instance, quantum machine learning for classical data as well as partial differential equations are not included  \cite{jiang2015quantum, arrazola2019quantum, larose2020robust, lubasch2020variational, xu2021variational, schuld2021effect, childs2021high, cerezo2022qml, amankwah2022quantum, jumade2023data, bravo2023variational}, because most implementations of such algorithms do not in fact first map the data to a qubit-based Hamiltonian. %
This is an important research area that will require its own curated high-quality datasets, and some work has already been done in this area \cite{Perrier21_qdataset}.

\subsection{Related work}\label{sec:related}

Benchmarking for classical computing has matured over many decades. Perhaps the most prominent large-scale linear algebra benchmark is LINPACK \cite{dongarra2003linpack,mohammadi2018linpack}, though widely used classical benchmarking standards exist, including MLPerf \cite{mattson2020mlperf}, SPEC \cite{dixit1991spec}, HPCG \cite{dongarra2013hpcg}, Green500 \cite{feng2007green500}, and Graph500 \cite{murphy2010graph500}. More analogous to the current work are past efforts constructing large \textit{datasets} that can be used in various benchmarking tasks. For instance, as mentioned above, ImageNet \cite{deng2009imagenet} contains millions of images intended for use in image analysis and deep learning, while similarly large datasets exist for simulating chemistry \cite{ramakrishnan2014qm9} and materials science \cite{jain2013matproj}.

In recent years there have been several proposals for benchmarking quantum algorithms and quantum computers. Perhaps most similar to previous classical benchmarking efforts are the proposed benchmark suites consisting of common quantum subroutines, such as quantum phase estimation and quantum adders \cite{parekh2016benchmarking,chen2022veriqbench,li2022qasmbench,Lubinski21_bench,cornelissen2021scalable,tomesh2022supermarq}. Benchmarks have been proposed specifically for near-term hybrid quantum-classical algorithms \cite{yeter2021benchmarking,mccaskey2019quantum,wu2023vscore}, with some benchmarking work concentrating on more specific topics, such as particular classical optimization problems \cite{Mesman21_qpack,finvzgar2022quark,lubinski2023optimization}, benchmarking the optimizer used in the classical step of hybrid algorithms \cite{mcclean2016theory,guerreschi2017practical,singh2022optimizes}, a dataset used for quantum machine learning \cite{Perrier21_qdataset}, and the dataset of curated Hamiltonians available in the software package Pennylane \cite{bergholm2018pennylane}. 
Narrower benchmarks have been proposed for very specific applications as well \cite{knill2000algorithmic,hammerer2005quantum,zhukov2019quantum,yeter2019scalar,eisert2020quantum,dallaire2020application,resch2021benchmarking,dong2021random,nakayama2023vqedataset}.
Other proposals include those related to quantum volume \cite{cross2019validating,magesan2012characterizing}, random benchmarking \cite{knill2008randomized}, and related techniques \cite{erhard2019cyclic,proctor2022mirorring}; such efforts are not associated with specific quantum algorithms but are an essential aspect of hardware characterization and design.

Despite the above-mentioned progress, there is not yet a large publicly available dataset of quantum Hamiltonians (i.e. cost functions) for benchmarking algorithmic implementations e.g. QPE, adiabatic quantum computing, VQE, and QAOA. The purpose of the current work is to fill this gap.

\subsection{Availability and dataset structure}

At the time of publication, the HamLib dataset is downloadable from the following URL:

\begin{center}
\url{https://portal.nersc.gov/cfs/m888/dcamps/hamlib/}
\end{center}
The Hamiltonians are organized into the following high-level categories:
\begin{itemize}
    \item \textbf{Binary-variable optimization and related problems}. Subcategories are Max-k-SAT, MaxCut, and QMaxCut.
    \item \textbf{Discrete-variable optimization problems} %
    (discrete variables taking $>2$ values). Subcategories are Max-$k$-Cut and the traveling salesperson problem.
    \item \textbf{Condensed matter} physics models commonly studied by theoretical physicists. Subcategories are the transverse-field Ising, Heisenberg, Fermi-Hubbard, and Bose-Hubbard models.
    \item \textbf{Chemistry} Hamiltonians that use curated or calculated real-world parameters. Subcategories are electronic structure and vibrational structure.
\end{itemize}

\begin{figure}
    \centering
    \includegraphics[width=\linewidth]{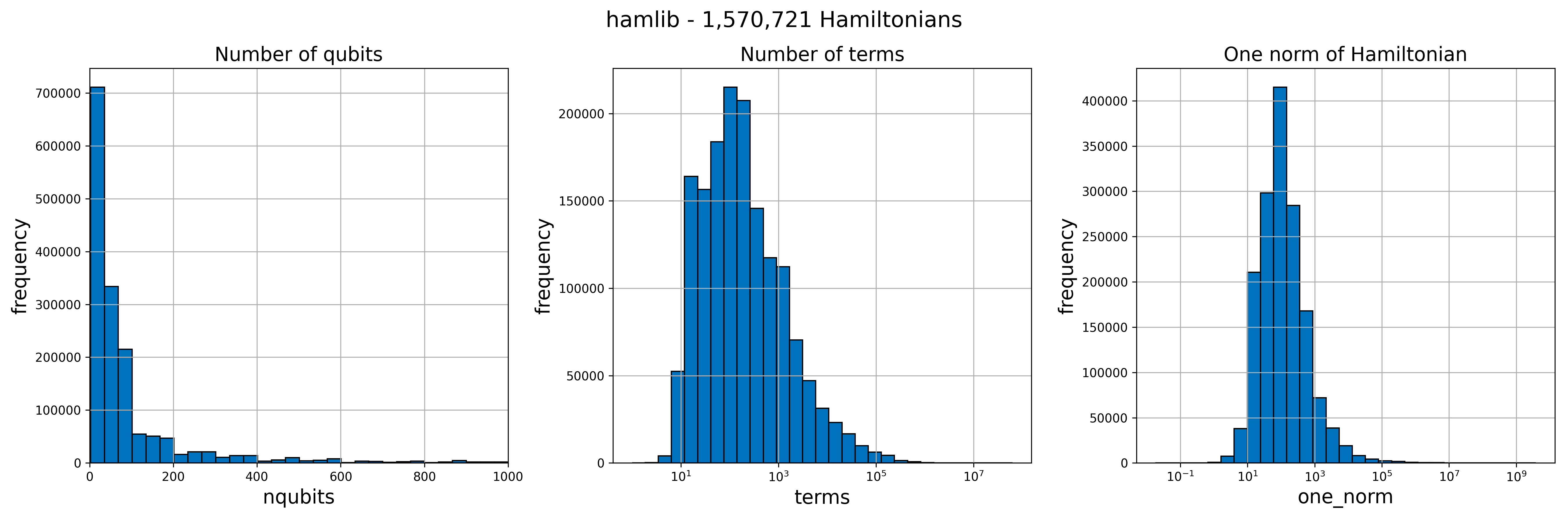}
    \caption{\textcolor{black}{Distribution of the number of qubits (\textbf{left}), number of terms (\textbf{middle}), and one norms (\textbf{right}) over all qubit Hamiltonians in HamLib.}}
    \label{fig:hamlib-distribution}
\end{figure}

As stated above, a major purpose of this work is to provide a large set of Hamiltonians that have already been mapped to a qubit representation. All encoded Hamiltonians are represented using OpenFermion's QubitOperator class~\cite{openfermion} and stored in the HDF5 format. These are compressed to ZIP format. \textit{Please note}, it will not be unusual for the decompression to yield an HDF5 file that is an order of magnitude larger than the ZIP file.
We provide a few Python code snippets useful to interact with the HamLib dataset \textcolor{black}{in the `\texttt{hamlib$\_$snippets.py}' file available via the URL linked above and summarized} in Appendix~\ref{app:snip}. \textcolor{black}{Figure~\ref{fig:hamlib-distribution} shows the distribution of the number of qubits, the number of terms, and the one-norms of the more than 1.5 million qubit Hamiltonians that are included in HamLib. Appendix~\ref{app:figures} includes similar plots for each of the four subdatasets.}

It is important to point out that for some of these Hamiltonians, minimization (e.g. minimum eigenvalue finding) is a typical goal, while for others maximization (e.g. maximum eigenvalue finding) is more natural.
We point this out this for any researcher considering running the entire dataset through a set of algorithms in a black-box manner.

In addition to the qubit representations of the operators herein, we include files for the original ``pre-encoded'' (before having been encoded into qubit representation) Hamiltonians where appropriate. This may take the format of a mat2qubit, OpenFermion, or SciPy array object as specified in the relevant sections of this document. Where appropriate, other auxiliary data is included, e.g. graph instances or approximate ground state values. If such auxiliary data is present for a given subset then this is mentioned in the relevant portion of Section \ref{sec:probinst}.

In Hamiltonians with non-binary and non-fermionic variables, the correspondence between the qubit indices and the original variable indices can be determined as follows. 

Smaller variables indices in the original variables space correspond to smaller variable indices in qubit space. \textcolor{black}{For example, if $y_0$ and $y_1$ are 8-valued variables encoded with standard binary, then qubit indices \{0,1,2\} correspond to $y_0$ and qubit indices \{3,4,5\} correspond to $y_1$. Further, within an encoded discrete variable the least significant figures correspond to the smaller qubit index. Hence $y_0 = q_0 2^0 + q_1 2^1 + q_2 2^2$, where $q_i$ is the value of the $i$th bit. 
In other words, we follow little-endian convention to store each individual variable and order variables in ascending order in qubit space.}

We provide an additional example for clarity. Assume we have variables $x_0$, $x_1$, and $x_2$, all with cardinalities 4 and encoded with standard binary. \textcolor{black}{The state $x_0=2 \rightarrow 10_2$, $x_1=1 \rightarrow 01_2$, and $x_2=0 \rightarrow 00_2$ (subscript denoting binary representation) corresponds to the quantum basis states \texttt{01-10-00}, where the hyphen \texttt{-} is included to visually separate the 3 variables. The same example in unary (one-hot) representation for which 
$x_0=2 \rightarrow 0100_2$, $x_1=1 \rightarrow 0010_2$, and $x_2=0 \rightarrow 0001_2$, becomes \texttt{0100-0010-0001}.}

\subsection{Modifications to and extensions of HamLib}

We believe that, for the purposes of this project, it is best to maintain \textit{static} datasets. This allows comparisons between research papers to be as valid as possible over the years. However, there are of course many other problem types that could have been included in this dataset, including e.g. those related to high energy physics or nuclear structure. Hence future modifications to HamLib will likely take the form of additional datasets, whether e.g. adding new distinct subsets of electronic structure Hamiltonians, or adding subsets for entirely new problem areas. Additionally there is some chance that some modifications will be made to the first version of the dataset. 

The \textcolor{black}{current} version of the dataset is denoted HamLib v1.1. If modifications are made to HamLib, we will assign a new version number to the changes, and include a full description of the changes or additions in the present section of this manuscript. HamLib versions:

\begin{itemize}
    \item \textbf{v1.0} - June 2023 - Initial release of HamLib.
    \item \textcolor{black}{\textbf{v1.1} - September 2024 - Updates:}
    \begin{itemize} 
    \item \textcolor{black}{All qubit Hamiltonians have the attributes `\texttt{nqubits}', `\texttt{terms}' and `\texttt{one$\_$norm}' as metadata attached in the HDF5 files.}
        \item \textcolor{black}{70- and 80-atom electronic structure hydrogen chains have been removed.}
    \item \textcolor{black}{All (sub)datasets include a CSV file with a list of all problem instances and their Hamiltonian statistics (number of qubits, number of terms, and one norm).}
    \item \textcolor{black}{All (sub)datasets include histograms showing the distribution of number of qubits, number of terms and one norms of the Hamiltonians.}
    \item \textcolor{black}{Added support for qiskit's \texttt{SparsePauliOp} Hamiltonian format.}
    \item \textcolor{black}{Minor fixes and improved file consistency across HamLib.}
    \end{itemize}
\end{itemize}

\section{Problem instances}\label{sec:probinst}

\subsection{Binary-variable optimization and related problems}

This section includes combinatorial problems defined over binary variables. We consider MaxCut, Max-$k$-SAT, and QMaxCut, a local Hamiltonian problem that is a quantum analogue of MaxCut.  We make a point of using problem instances from established libraries used in classical benchmarking, %
as well as random graphs we constructed for this work. For the former classes of graphs, where appropriate, we have implemented \textit{variable reduction} schemes whereby smaller problem instances are created from the large original instances~\cite{Hastad2002randomassignment,dupont2006subgraphextraction,lokshtanov2012kernelization}. The purpose of these problem reductions is to ensure that the qubits counts are well-spaced as discussed in Section \ref{sec:attributes}, while retaining some of the structure of the original problem instance.

\subsubsection{MaxCut}\label{sec:maxcut}

The prototypical classical problem to which NISQ algorithms (especially QAOA) are applied is the MaxCut problem, for which the cost function is defined as
\begin{equation}
H_C = \sum_{(j,k)\in \mathcal E} \half \left( 1 - Z_j Z_k \right),
\end{equation}
where $\mathcal E$ is the set of edges in the graph $\mathcal G$.
MaxCut is a commonly studied graph partitioning problem, which in turn has applications in many domains. In addition, MaxCut (the optimization version) is a classic NP-hard problem. %

In HamLib we include a set of trivially 2-colorable graphs, which may be useful for sanity checks---\textit{i.e.} if a quantum algorithm cannot find the optimum for these trivial graphs then it may indicate a pathology in the algorithm. These are the complete bipartite graphs (with partition sizes $a$ and $b$), star graphs, and circulant graphs.

We constructed two general types of random graphs. First are random $X$-regular graphs for regularities 3,4,5,6. Additionally, we include Erdos-Renyi graphs GNP$(n,p)$, where $p$ is the probability that a given edge is present. Though previous work in quantum algorithms has studied the case of GNP$(n,\frac{d}{n-1})$ for fixed parameter $d$ \cite{farhi2020qaoagnp,boulebnane2021qaoagnp}, we have chosen not to use this graph class because nearly all such graphs are disconnected \cite{erdos1960evolution}. We instead implement GNP$(n,k\frac{\ln n}{n})$ for fixed parameter $k$, which for $k>1.0$ is virtually guaranteed to be a connected graph \cite{erdos1960evolution}. For these constructed graphs we include variable counts of: 4 to 10 (step 1); 10 to 48 (step 2); 50 to 200 (step 10); 300 to 1000 (step 100). These instances are used in Max-$k$-Cut and QMaxCut as well.

Finally, we derive graphs from Biq Mac \cite{wiegele2007biqmac}, a library that is widely used in classical benchmarking, and CI-QuBe \cite{CI-QuBe2021}, a collection of combinatorial instances for quantum benchmarking. Biq Mac contains two classes of instances: the Ising instances are lattice graphs taken from statistical physics and the `rudy' instances are generated from the rudy graph generator.
From CI-QuBe we used the Karloff \cite{karloff1996good} and ratio912 data sets. 
Karloff shows that there is a sequence of graphs whose Goemans-Williamson (GW) approximation ratio approaches 0.878; these are the graphs at the beginning of such a sequence using the construction Karloff proposes. 
The ratio912 graphs are graphs in the MQLIB instance library \cite{dunning2018works} that have a 0.912 approximation ratio with respect to GW. For these instances, the optimal cut is always 2/3 of the total number of edges. These GW approximations provide some insight into the hardness of each dataset.

The goal of the problem reductions is to preserve much of the structure of the original Biq Mac and CI-QuBe graphs, ensure the resulting reduced graphs are connected, and ensure that the instances are well-spaced in terms of qubit counts. To this end, we use two variable reduction methods to generate instances with arbitrary node counts. The first is random restriction \cite{Hastad2002randomassignment,jarvisalo2008randomrestr,lokshtanov2012kernelization} where we randomly partition a sample of vertices (i.e. we reduce the graph by removing already partitioned vertices). The second is a random walk \cite{dupont2006subgraphextraction} method that creates a randomly connected sub-graph. We initially attempt to use random restriction, but in cases where the random restriction is unable to create a sufficient number of connected instances, we use the random walk method.

The following table summarizes the graphs used for the HamLib instances of MaxCut.
\\

\begin{tabular}{ | p{0.15\linewidth} | p{0.75\linewidth} 
 |}
\hline
 Trivial graphs & 2-colorable graphs: Complete bipartite, star, circulant with offsets \{1,2\}. \\  
 \hline
 Random & $X$-Reg with $X \in \{3,4,5,6\}$, and GNP$(n,k\frac{\ln n}{n})$%
 with $k \in \{ 2,3,4,5 \}$. \\   \hline
  Biq Mac lib & Variable reduction method:  random restriction and random walk. (Ising and Rudy) \\ 
 \hline
  CI-QuBe \cite{CI-QuBe2021} & Variable reduction method: random restriction and random walk. (Karloff and ratio912) \\ 
  \hline
  \hline
  (Auxiliary) & Graph instances \\

\hline
\end{tabular}

\subsubsection{Max-k-SAT}

In addition to being relevant to industrial optimization, satisfiability problems form the basis for much of theoretical computer science. Specifically, 3-SAT is often used for complexity theoretic proofs \cite{arora2009complexitytext}, partly because any NP-Hard problem can be cast as a 3-SAT problem.

Constructing a Hamiltonian to represent a 3-SAT problem involves summing 3-variable terms such as

\begin{equation}
\begin{split}
x_i \lor x_j \lor x_k = I - \frac{1}{8} (I+Z_i)(I+Z_j)(I+Z_k)
\end{split}
\end{equation}
if negations are not present. When negations are included (as they are in HamLib), the expression is
\begin{equation}\label{eq:3satclause}
\begin{split}
(\lnot)^{s_i}x_i \lor (\lnot)^{s_j}x_j \lor (\lnot)^{s_k}x_k = I - \frac{1}{8} [I+(-1)^{s_i}Z_i][I+(-1)^{s_j}Z_j][I+(-1)^{s_k}Z_k]
\end{split}
\end{equation}
where $s_j=1$ if $x_j$ is negated in the clause, else 0.

From classical computer science, it is known that the hardness of solving satisfiability problems increases with the clause ratio $r=m/n$, where $m$ is the number of clauses and $n$ is the number of variables. Analytical and theoretical results exist which estimate or place bounds on the ratio $r$ at which a problem becomes intractable \cite{achlioptas2005nature}. In quantum computation as well, researchers have studied hardness scaling in random SAT instances \cite{akshay2020reachability,zhang2022ksat}.

Here, we choose clause ratios ($r \in \{2,3,4,5\}$) that cross the (both numerically and analytically estimated) thresholds for hardness in both classical \cite{achlioptas2005nature} and quantum computational studies \cite{akshay2020reachability,zhang2022ksat}.
Implementing variable (qubit) counts $n$ from 4 to 1000, for each $(n,r)$ we create 10 random Hamiltonian instances.  

In addition to generating random instances, we also pull a sample collection of datasets from SATLib \cite{hoos2000satlib}. We use the following datasets: aim, lran, flat30, flat100, uf20-20, uf100-430, uf200-860, uf250-1065, uuf100-430, uuf200-860, and uuf250-1065. SATLib is a representative library of Max-$k$-SAT instances that contains many satisfiable and unsatisfiable instances. For variable reduction, we first convert the clauses into a graph where each vertex represents a clause, and an edge exists between two vertices if the associated clauses share a variable. Then we apply random restriction \cite{jarvisalo2008randomrestr} to the variables and give random assignments to a random sampling of variables. As we remove variables and clauses, we update the graph representation accordingly. 
If the graph is disconnected, we compute the connected components and then greedily connect them by adding additional clauses.

Note we include the raw pre-encoded SAT instances as lists of 1-indexed integers, while the qubit operators are 0-indexed. We implement the following classes of instances for Max-k-SAT.
\\

\begin{tabular}{ | p{0.1\linewidth} | p{0.8\linewidth} 
 |}
 \hline
 Random instances & Clause ratios $r \in \{2,3,4,5\}$ \\  
 \hline
 SATLib & Limiting to 2- and 3-SAT. Random restriction (10 random instances per problem size) to reduce problem size. \\ 
   \hline \hline
  (Auxiliary) & SAT instances\\
\hline
\end{tabular}

\subsubsection{QMaxCut}

Here we consider a ``quantum version'' of MaxCut. %
Though this problem is structurally similar to its classical counterparts, its ``quantumness'' leads to rich and surprising differences in computational complexity with respect to its classical counterpart \cite{parekh2023synergies}.

Quantum MaxCut (QMaxCut) is a quantum optimization problem (i.e., a local Hamiltonian problem) that can be understood as a generalization of the MaxCut problem.
Given a (possibly weighted) graph \( \mathcal{G} \) with vertices \( i \in \mathcal{V}\), edges \((i,j) \in \mathcal{E}\), and non-negative weights \( w_{ij} \in \mathcal{W} \), the problem can be defined as finding the \textit{maximum} eigenvalue of the following Hamiltonian:
\begin{equation}
    H = \sum_{(i,j) \in \mathcal{E}} w_{ij} H_{ij}, 
\end{equation}
where \(H_{ij} = \frac{1}{4} \left( I - X_i X_j - Y_i Y_j - Z_i Z_j \right)\). All instances in HamLib use $w_{ij} \in \{0,1\}$. 

In general, QMaxCut is QMA-Hard \cite{piddock2015complexity}, where QMA is the quantum analog of the NP complexity class in classical computing \cite{papadimitriou2003computational}.
There are several approximation algorithms for this problem, with the best approximation ratio for generic instances being 0.562 \cite{lee2022optimizing}.

This problem is quantum in the sense that instead of finding the maximum eigenvalue of a real-coefficient degree-2 polynomial \(P\) over commutative variables \(I, Z_1, \dots, Z_n\), the problem is to find the maximum eigenvalue of a real-coefficient degree-2 polynomial \(Q\) over non-commutative variables \(I, X_1, Y_1, Z_1, \dots, X_n, Y_n, Z_n\).
This polynomial can be expressed as a Hermitian matrix \(Q \in \mathbb{C}^{2^n \times 2^n}\) instead of a diagonal matrix \(P \in \mathbb{C}^{2^n \times 2^n}\).  

QMaxCut resembles MaxCut in that each term $H_{ij}$ is a scaled version of the corresponding MaxCut term of $H_C$ with additional Pauli $X$ and $Y$ terms.  In particular, the diagonal of $H$ above is a scaled version of the MaxCut Hamiltonian, $H_C$.  These structural similarities have enabled generalizations of approximation algorithms and hardness results for MaxCut to the QMaxCut setting~\cite{hwang2023unique}.  In this sense, QMaxCut has served as a testbed for developing algorithmic ideas to approximate quantum local Hamiltonians, just as MaxCut serves as a canonical classical constraint satisfaction problem.   

Finding a maximum energy state for QMaxCut is an equivalent problem to finding a minimum energy state of a quantum Heisenberg model, described below; although this leads to subtle differences from the point of view of approximate solving \cite{gharibian_et_al:LIPIcs:2019:11246}.
This is one instance of what is known as quantum 2-local Hamiltonians, for which finding their minimum or maximum eigenstates is relevant for quantum many-body physics.

While MaxCut has been a focal point for quantum algorithms such as QAOA, it is not clear whether quantum algorithms can offer approximation advantages for classical problems.  For example, it is NP-hard to approximate MaxCut better than the celebrated Goemans-Williamson algorithm~\cite{goemans1995improved} under the Unique Games Conjecture~\cite{khot2007optimal}, and NP-hardness is likely a barrier for polynomial-time quantum algorithms as well.  Thus it may be fruitful to consider inherently \emph{quantum} problems when seeking quantum approximation advantages.  Yet the types of local Hamiltonian problems typically studied in physics have not generally been amenable to approximation algorithms in the same way that classical problems have been.  We include QMaxCut in HamLib because it may serve as a canonical quantum problem in the study of quantum approximation advantages.  Approximation algorithms for QMaxCut employ techniques for approximating MaxCut as well as quantum approaches such as variational algorithms~\cite{gharibian_et_al:LIPIcs:2019:11246,anshu_et_al:LIPIcs:2020:12066,parekh_et_al:LIPIcs.ICALP.2021.102,parekh2022optimal,king2022improved,lee2022optimizing}.  HamLib's problem instances of QMaxCut use the same graph instances as MaxCut. The relevant table of instances is in Section \ref{sec:maxcut}.

\subsection{Discrete-variable ($d>2$) optimization problems}

Here we consider classical cost functions that are defined over discrete variables taking more than 2 values. As in the case of binary optimization problems, we create both random instances and instances derived from widely-used libraries.

Unlike MaxCut and Max-$k$-SAT, there is a rich set of choices for how these non-binary variables are encoded into qubits \cite{sawaya2022dqir}. We implement the unary (i.e. one-hot), Gray, and standard binary encodings for the discrete combinatorial problems of this section. A notable omission is the domain wall encoding, which has been shown to improve computational efficiency in some cases \cite{chancellor2019domainwall}; we did not include it because this encoding is not yet included in mat2qubit~\cite{mat2qubit}. Note that different encodings require different numbers of qubits per variable, which in turn can lead to space-depth tradeoffs when choosing encodings \cite{sawaya2020resource,sawaya2022dqir}.

\subsubsection{Max-k-Cut}

The Max-k-Cut problem is a generalization of MaxCut. For a graph $G = (V, E)$ with weight function $w: E \to \mathbb{R}$ for the edge set $E$, the Max-k-Cut problem amounts to finding a partition of $V$ into $k$ disjoint sets $\lbrace C_1, \ldots, C_k \rbrace$ that maximizes the sum of the edges between disjoint sets, i.e., the cost function is
\begin{equation}
\sum_{\substack{i,j \in [k]\\ i > j}} \sum_{\substack{u \in C_i\\ v \in C_j}} w(\lbrace u, v \rbrace).
\end{equation}

In this work, we consider only unweighted graphs such that the weight function becomes the indicator function $\mathbf{1}_E$, i.e., 1 if $\lbrace u, v \rbrace \in E$ and 0 otherwise. This unweighted version of \maxkcut\ is equivalent to finding the maximum k-colorable subgraph. \maxkcut\ is APX-complete~\cite{Frieze1997} and is equivalent to MaxCut for $k = 2$.

We implement the same random graphs created for the MaxCut section, as well as a set of instances derived from COLOR02~\cite{color02}, a graph library used for benchmarking, including classical studies of Max-k-Cut~\cite{Galinier2013, LUCET20062189}. In the latter case, we use a random depth-first search \cite{dupont2006subgraphextraction} to achieve arbitrary-sized, connected subgraphs of the original problem. For all problems in the Max-k-Cut dataset, we use three encodings: unary (i.e. one-hot), Gray, and standard binary.
We summarize the Max-k-Cut instances in the following table. We use $k \in \lbrace 3,4,5 \rbrace$.
\\

\begin{tabular}{ | p{0.1\linewidth} | p{0.8\linewidth} 
 |}
  \hline
 Trivial graphs  & Complete $k$-partite with $k \in \lbrace 3, 4, 5 \rbrace$, circulant with offsets $\lbrace 3, 4, 5, 6 \rbrace$, windmill with $k \in \lbrace 3, 4, 5, 6 \rbrace$ cliques.\\
 \hline
 Random & Random $X$-Regular for $X \in \lbrace 3, 4, 5, 6 \rbrace$, GNP$(n,d/n-1)$ for $d \in \lbrace 2, 3,4, 5, 6 \rbrace$. \\
\hline
 COLOR02 & Variable counts reduced via random depth-first search (10 random restrictions per graph) up to 200 nodes. \\
 \hline
 \hline
 (Auxiliary) & Graph instances \\
\hline
\end{tabular}

\subsubsection{Traveling Salesperson Problem}

The traveling salesperson problem \cite{burkard1998tspsolvable,laporte1992tspalgos} (TSP) is a widely studied problem in combinatorics, and relates to routing problems encountered in industry. Our dataset includes both cost Hamiltonians and permutation penalties for the traveling salesperson problem, whose cost function is defined as 
\begin{equation}
\sum_{a=0}^{M-1} d \left( \tau(a),\tau(a+1 \mod M) \right).
\end{equation}

where $\tau$ is a permutation over a set of $M$ cities. We encode this problem as a permutation of integers (as opposed to a binary problem with constraints). One disjoint set of qubits is used per city, \textit{i.e.} if there are $N_{c}$ cities, then $N_{c} \times \m D(N_{c})$ qubits are required, where $\m D(K)$ is the number of qubits required to encode a discrete variable of cardinality $K$. This way of representing TSP is amenable to encoding-dependent penalties that have previously been called \textit{pair permutation penalties} \cite{sawaya2022dqir}.%
When implementing this problem, one may alternatively use penalty-free methods that constrain the state \cite{hadfield2019qaoa}, including graph-derived partial mixers \cite{sawaya2022dqir}.

Notably, for the decision version of TSP, there is a phase transition \cite{gent1996tspphasetr} with respect to parameter $\alpha=l/\sqrt{nA}$, where $n$ is the number of cities, $A$ is the area, and the question is whether a path length of $\leq l$ exists. When $\alpha<\alpha^*$ there is unlikely to be any path shorter than $l$. This phase transition has been observed to occur approximately at $\alpha^*=0.78$. We use this relation to guide our construction of random instances. For a given $n$, we select random positions in a square of area $A=l^2/(\alpha^*)^2 n$. Roughly speaking, this implies that there is approximately a 50\% chance that a path of length $l$ exists; we arbitrarily choose $l=100$.

For the 77 problem instances in TSPlib \cite{reinelt1991tsplib} containing fewer than 1000 cities, we reduce the number of cities one at a time, by removing the city with the lowest weight, \textit{i.e.} the city that has the shortest distance to its two closest neighbors. The justification is that the two shortest distances are likely to be connected in the optimal route; hence the rest of the path is unlikely to radically change. This is just a heuristic and we acknowledge there may be topologies for which removing cities in this way will qualitatively change the solution to the problem. 
However, we note that this variable reduction procedure is conceptually similar to what is done in previously used greedy algorithms \cite{geng2011tspgreedy,laporte1992tspalgos}. %

We implement the following problem instances of TSP:

\begin{tabular}{ | p{0.15\linewidth} | p{0.75\linewidth} 
 |}
 \hline
 Random & Random points chosen in a square of area  $A=100^2/0.78^2 n$. \\  
 \hline
 TSPLib & All symmetric instances from TSPLib with $N_{cities}<1000$. Problem size reduction is implemented by iteratively removing the node with the smallest weight. \\ 
\hline
\hline
 (Auxiliary) & Distance matrices for all problem instances. \\
\hline
\end{tabular}

\subsection{Condensed matter physics models}

Here we describe the condensed matter models included in the dataset. We implement spin, fermion, and boson models, covering 1D, 2D, and 3D lattices. We expect the fermionic and bosonic models to be the more useful contributions of this section as they involve more effort to prepare.

We use both periodic boundary conditions (PBC) and non-PBC for all lattices. For users who would like to add local disorder, we have included some qubit-encoded occupation number operators for each individual site; these may be used to add local disorder.

\subsubsection*{Lattices}

We include 1D grids, 2D grids, 2D ribbons, and 3D grids. The 2D instances include square, triagonal, and hexagonal (honeycomb) arrangements. Qubit counts of 2 through 1000 are used, with spacing increasingly sparse as qubit count increases. %
Importantly, we implement a ``snaking'' pattern for all grids, which leads to more efficient fermionic representations \cite{verstraete2005mapping}. We include the mapping from node ID to grid position as well, which may optionally be obtained using the \texttt{read\_gridpositions\_hdf5()} function found in the Appendix.\\

\begin{tabular}{ | p{0.25\linewidth} | p{0.65\linewidth}  |}
\hline
 1D & PBC and non-PBC. \\ 
 \hline
 2D & Square, triangle, hexagonal. PBC and non-PBC. \\  
  \hline
 2D ribbon & Width 2 through 5 for square lattice. PBC and non-PBC. \\  
   \hline
 3D & Cubic lattices. PBC and non-PBC. \\  
    \hline
    \hline
 (Auxiliary) & Grid positions (mapping of node IDs to positions). \\  
 \hline
\end{tabular}

\subsubsection{Transverse-field Ising model}

The transverse-field Ising model (TFIM) is the simplest of the condensed matter models that we include. Notably, the one-dimensional version of TFIM is classically tractable even in the presence of disorder. The Hamiltonian is defined as
\begin{equation}\label{eq:tfim}
H = \sum_i h_i X_i + \sum_{\la i,j \ra} Z_i Z_{j}, %
\end{equation}
where the sum is over each edge $\la i,j \ra$ in the lattice. 
Quantum critical points have been found to exist at $h \approx 3$ in two-dimensional models \href{https://journals.aps.org/prl/abstract/10.1103/PhysRevLett.110.135702}{[Kallin et al 2013]}, 
as well as at $h \approx 5.16$ in three-dimensional TFIM \href{https://journals.aps.org/prresearch/abstract/10.1103/PhysRevResearch.3.023236}{[Tepaske and Luitz 2021]}. 
These values guide our choices for HamLib---in order to include Hamiltonians on both sides of these critical points we implement values $h \in \{0,0.1,0.5,1,2,3,4,5,6\}$ \cite{blote2002tfimcluster}.

\subsubsection{Heisenberg model}

We implement the following Hamiltonian for the quantum Heisenberg model, 
\begin{equation}
H_{\textrm{Heis}} = \sum_{i=1}^N \left( \vec \sigma_i \cdot \vec \sigma_{i+1} + h_i Z \right).
\end{equation}
where $\vec \sigma_i = (X_i,Y_i,Z_i)$.
The Hamiltonian is also known as the Heisenberg XXX model with external magnetic field and can be solved by the Bethe ansatz in the one-dimensional case without disorder~\cite{Franchini2017,Granet2019}.
Based on values used in previous work, HamLib includes $h \in \{0,0.1,0.5,1,2,3,5\}$ \cite{childs2018toward,akinci2013critical}. %
While HamLib only includes the Heisenberg XXX model, anisotropic variants such as the Heisenberg XXZ and XYZ models \cite{Werlang81XYZ} can easily be generated from the lattice data.

\subsubsection{Fermi-Hubbard model}

The Fermi-Hubbard Hamiltonian \cite{hubbard1963electron} captures the behavior of fermions on lattice sites. It is defined as
\begin{equation}
    H_{FH} = -t\sum_{\langle i,j\rangle,\sigma}\big(c_{i,\sigma}^\dagger c_{j,\sigma}+c_{j,\sigma}^\dagger c_{i,\sigma}\big)+U\sum_i n_{i\uparrow}n_{i\downarrow},
    \label{Eq:FermiHubbardHamiltonian}
\end{equation}
where $\langle i,j\rangle$ labels adjacent lattice sites $i$ and $j$ , $\sigma$ labels the fermion spin, $c$ and $c^\dagger$ are the fermionic annihilation and creation operators, respectively, and $n_{j\sigma} = c_{j\sigma}^\dagger c_{j\sigma}$ is the number operator associated with spin $\sigma$ and site $j$. The first term contains the noninteracting portion of the Hamiltonian and describes fermions hopping between adjacent sites with tunneling amplitude $t$. The second term describes the on-site interaction between fermions with strength $U$. The Fermi-Hubbard model can be solved analytically for $U=0$ and $t=0$. However, in the general case, an analytical solution is only known in 1D \cite{lieb2003one}. In 2D, the Fermi-Hubbard model has been investigated extensively through simulations, with findings suggesting that the 2D Fermi-Hubbard Hamiltonian in the intermediate coupling regime ($U/t = 4,6,8$) near half-filling is especially difficult to solve \cite{leblanc2015solutions}. %

HamLib includes numerous instances of the Fermi-Hubbard Hamiltonian for different lattice structures and dimensions. We implement three fermion-to-qubit mappings for each instance: Jordan-Wigner, parity, Bravyi-Kitaev \cite{cao2019chemrev}. We include files for the pre-encoded fermionic encodings as well. Based on previous studies on 1D, 2D, and 3D versions of the Fermi-Hubbard model \cite{leblanc2015solutions}, the parameters we use are the same across all grid classes. In particular, we take $t=1$ and $U \in \{0,2,4,6,8,12\}$. We include pre-encoded fermionic Hamiltonians in OpenFermion format.

\subsubsection{Bose-Hubbard model}

The Bose-Hubbard (BH) model is defined as 
\begin{equation}\label{eq:bh}
H_{BH} = - t\sum_i \left(  b_{i+1}^\dag b_i + h.c.  \right) + \frac{U}{2} \sum_i n_i (n_i - 1)
\end{equation}
where $b^\dag_{i}$ ($b_i$) are creation (annihilation) operators, the number operator $n_i \equiv b^\dag_i b_i$, $t$ is the tunneling strength (set to $t=1$ in this work), and $U$ is the site energy. Equation \eqref{eq:bh} often includes an additional term proportional to the chemical potential $\mu$, which effectively sets the particle count. We omit this chemical potential term in our dataset, with the assumption that the user will set an arbitrary number of particles at the beginning of the simulation.

The standard version of this model exhibits two phases, a Mott insulator and a superfluid \cite{fisher89bh,freericks1994bh}. More complex versions of the model lead to more exotic phases including density wave \cite{pai2005bh_densitywave} and supersolid phases \cite{batrouni95bh_supersolid}. There has been substantial previous work on qubit-based simulations of bosonic degrees of freedom \cite{somma2005quantum,macridin2018pra,macridin2018prl,klco2019scalarfields,sawaya2020resource,sawaya2020connectivity,tong2021provably,liu2021towards,bahrami2024bempa}, though to our knowledge no qubit-based experimental demonstrations to date.

For our dataset, we chose dimensional-dependent values for $U$, where parameters were based on the phase diagrams reported in reference \cite{freericks1994bh}. The transition between Mott insulator and superfluid is dependent on the filling $\langle n_0 \rangle$, the average number of particles per site. The user may effectively set $\langle n_0 \rangle$ during simulation (\textit{e.g.} by choosing a particular product state at the beginning of the simulation). For most values of $\langle n_0 \rangle$ (which may be set for example by choosing a Fock state at the beginning of the simulation), this range for $U$ will yield some Hamiltonians in the Mott insulator regime and some in the superfluid regime.
For 1D, $U/t \in [2,10,20,30,40]$. For 2D, $U/t \in [10,30,50,70,100]$. For 3D, $U/t \in [20,40,60,90,120]$.

We implement 3 mappings for each instance: unary (one-hot), standard binary, and Gray, where we implement bosonic mode truncations of both 4 and 8 ($d=4,8$). We included files for the pre-encoded bosonic encodings as well, in mat2qubit format.

\subsection{Chemistry}

\subsubsection{Electronic structure}

The molecular electronic structure Hamiltonian may be expressed as
\begin{equation}
H = \sum_{pq} h_{pq} a_p^\dag a_q + \half \sum_{pqrs} h_{pqrs} a_p^\dag a_q^\dag a_r a_s
\end{equation}
where $a^\dag$ and $a$ are the usual fermionic creation and annihilation operators. 
HamLib included molecules from a range of chemical classes; further, we ensured that the molecules include a range of difficulty. Unless otherwise specified, the nuclear coordinates used are those experimental geometries found in the online database of the National Institute of Standards and Technology (NIST), \url{https://cccbdb.nist.gov/geometries.asp}. \textit{Important:} for most molecules provided, some low-energy orbitals were frozen. The number of frozen orbitals for each molecule is given in Appendix \ref{sec:frozen_elec}. One must know the number of frozen orbitals in order to properly choose the correct number of effective electrons to use when implementing a quantum algorithm.

An important consideration was the choice of active space, as one goal of HamLib is to have problems with evenly-spaced qubit counts. Hence for each molecule, we include Hamiltonians of \textit{several qubit counts}, in order to allow the user to choose a level of complexity. We chose the smallest qubit count by calculating the CISD orbital occupation and keeping all orbitals for which the occupation is greater than 0.1. We use the def2SVP basis set for molecules containing transition metals, STO-6G for elemental hydrogen, and ccPVDZ for all others. We ensured that the coupled cluster CCSD(T) energy converges with this occupation condition. Then, to create several instances with more qubits, we then added orbitals to this set either one orbital at a time or one set of degenerate orbitals at a time, to create the 5 smallest qubit counts in addition to the large Hamiltonian that uses the full basis set.

Further, because the orbitals are ordered by energy, \textit{a user may straightforwardly create their own electronic structure Hamiltonian of arbitrary qubit count by removing all qubits with index above some number,} thereby choosing their own active space. This may be done with the \texttt{remove\_qindices()} function that we provide.
Our Hamiltonian sizes range up to 120 qubits, with the largest instances being hydrogen chains. %

The following molecules were picked based on different benchmark studies that have been performed in the chemistry literature in recent years. There are many high-accuracy classical methods for simulating such systems, and a recent study on benzene \cite{eriksen2020ground} highlighted many of the most modern methods that can be used for the same molecule. Studies like these are seldom done for a wide range of molecules, and thus we rely on many different papers that span a range of different molecules to look at.  %
Our dataset includes main group diatomics, hydrogen chains, and transition metal containing diatomics.  

For the transition metal molecules, we selected 10 systems that range from ``easy'' to ``hard.''
For these molecules, a recent work took a selected CI approach to use for a benchmark set of systems, and then compared different coupled cluster techniques against these results \cite{hait2019levels}.  We observed that there is no obvious standard ordering of the hardness of these systems that is consistent with all the different coupled cluster techniques.  For purposes of this work, we use CCSD(T) as the method in which to assess the difficulty of the systems, as it is one of the most widely used rigorous methods in quantum chemistry for treating correlated systems. By this definition, these molecules in order of difficulty are VH (1.0), ScO (1.4), ScC (3.9), TiH (4.9), CrO (8.1), MnN (9.6), FeC (10.7), CuC (12.3), CoH , (22.7), NiO (38.2), where in parenthesis is the CCSD(T) error in units of kJ/mol. We include the famously difficult chromium dimer (Cr\st) as well.

We now discuss Hamiltonians composed only of hydrogen. If quantum computers are going to be used for computing properties of molecules and materials, they must focus on the most challenging Hamiltonians for which classical algorithms struggle. In electronic structure, these are systems with a large number of open-shell (singly occupied) orbitals, which leads to strong electron correlation among the valence electrons.\cite{Lee2023, Ollitrault2024, Marti-Dafcik2024b}

Examples of molecular systems with such features include stretched chemical bonds\cite{Chan2024, Marti-Dafcik2024a} or transition metal clusters.\cite{reiher2017pnas, Lee2023} However, in such systems, only a subset of electrons are strongly correlated and many of the orbitals/electrons are weakly correlated and can be approximated at the mean-field level. Thus, benchmarking quantum algorithms on a system with many strongly correlated electrons requires either using a large number of qubits or carefully selecting an active space that contains only the strongly correlated electrons. The selection of active spaces is a non-trivial procedure that generally requires a significant amount of chemical understanding of each system considered.\cite{li2019electronic}

Instead, here we provide a set of hydrogen clusters at different geometries and in a minimal basis.
They are a good testbed for benchmarking quantum algorithms because, at stretched geometries,
all their electrons are strongly correlated.
By arranging the atoms in different geometries and dimensions, these systems can be used for testing the ability of different wave function parameterizations to capture different entanglement regimes.

Hydrogen model systems have long served as a benchmark for electronic structure algorithms on classical computers \cite{Jankowski1980a, Motta2017, Stair2020a}, and are currently being used to test quantum or hybrid quantum-classical algorithms \cite{Grimsley2019a, Stair2020a, Klymko2022, Burton2023}.

For all four structural motifs (linear chains, rings, sheets and pyramids), we included $H_n$ systems with $n=10, 12, 14, 16$, using the STO-6G basis set. These were studied in Stair \textit{et al.} \cite{Stair2020b}, which benchmarked classical electronic structure methods including selected configuration interaction (selected CI), singular value decomposition full CI, and density matrix renormalization group (DMRG). This allows a direct comparison of any quantum algorithms with the classical benchmark data in said paper. For $n \leq 40$ we include systems with bond lengths ranging from $\SI{0.5}{\angstrom}$ to $\SI{2.0}{\angstrom}$ in steps of $\SI{0.1}{\angstrom}$; for $n > 40$ we include only the bond length 1 \AA.
For the linear $H_n$ structures, we include the atom counts $n=$ 2,4,6,8,10,12,14,16,18,20,  24,28,32,36,40,  50,60. %

For the systems with over 16 atoms, exact classical calculations are difficult or unfeasible and we therefore only provide the restricted Hartree-Fock (RHF) energy. The larger Hamiltonians may be of use for estimating the resources required for Hamiltonian simulation algorithms \cite{lee2021even}.

 We used 3 mappings for each problem instance: Jordan-Wigner, parity mapping, and Bravyi-Kitaev.
 We implement the following molecules for electronic structure.
 \\

\begin{tabular}{ | p{0.25\linewidth} | p{0.65\linewidth}  |}
\hline
 Hydrogen (H$_n$) & 1D chains, 1D rings, 2D sheets, 3D pyramidal clusters \\ 
\hline
 Main group diatomics & H\st, N\st, O\st, F\st, B\st, C\st, Be\st; LiH, BeH, BH, CH, NH, OH, HF, O$_3$, Li\st, NaLi, Na\st \\  
\hline
Trans. metals & VH, ScO, ScC, TiH, CrO, MnN, FeC, CuC (low spin), CoH, NiO; Cr\st \\  
\hline
 Bond breaking & F$_{2}$ (single bond),  O$_{2}$ (double bond), N$_{2}$ (triple bond) \\  
\hline
\end{tabular}

\subsubsection{Vibrational structure}

Molecular vibrational structure \cite{wilson1980bookvibr} is a problem that is often intractable on classical computers, especially when strong anharmonicity and resonanaces (e.g. Fermi resonances) are present \cite{jankowski2012ch5,mathea2021vci}. Accurate calculations of the vibrational structure of molecules are crucial in identifying unknown molecules in fundamental chemical physics experiments and in astrochemical settings. Confirmed assignments in vibrational spectroscopy also provide direct probes of potential energy surfaces of molecules and fundamental probes of reactivity. Further, the precise locations of vibrational energy levels, which are difficult to compute classically, can directly impact the kinetics of chemical reactions.

In the harmonic basis, the vibrational structure Hamiltonian is
\begin{equation}\label{eq:ham_aharm}
H = \half \sum_i^M \omega_{i}(q_i^2 + p_i^2) + \sum_{\{ijk\}} h_{ijk} q_iq_jq_k   + \sum_{\{ijkl\}} h_{ijkl} q_iq_jq_kq_l + \cdots,
\end{equation}
where $q_i$ are canonical position operators, $p_i$ are momentum operators, $\omega_i$ are the harmonic frequencies, and $h_{ij\cdots}$ are higher-order coupling constants. 

In some respects, vibrational structure might often be easier to implement on a quantum computer than electronic structure because the former does not require particle conservation and some properties of vibrational Hamiltonians may be more favorable \cite{sawaya2021ir}. Recent efforts in developing quantum algorithms for calculating vibrational structure have demonstrated how some vibrational classical algorithms (e.g., vibrational coupled cluster) can be adapted to quantum hardware \cite{mcardle2019vibr,sawaya2019vibronic,ollitrault2020hardware}. %

For this benchmarking set, our focus is on diversifying the set of Hamiltonians that exist for benchmarking new methods in vibrational structure. We include a limited set of diatomic molecules, but our primary focus is on triatomic and larger systems. Triatomic systems are the smallest systems that can have resonances that complicate their vibrational structures. 
Expanding to tetra-atomic and larger systems allows one to study interesting and nontrivial differences in physical behavior between related molecules. %
For instance, BH$_3$, BF$_3$, and BHF$_2$ each have a similar trigonal planar geometry, but their vibrational Hamiltonians are substantially different in practice due to the different mass distributions across the geometry \cite{schwartz1970ab}. 

We include a mix of molecule sizes and a mix of vibrational truncation levels, all of which are given in the key names of the HDF5 files. %
We include up to fourth-order terms excluding those with more than two unique indices (as such terms are often negligible), which is a common approximation in the quantum chemistry literature. %
This truncated expansion is common in vibrational structure calculations and with a good choice of the coordinate system is often well-converged. The force fields were generated with the CFOUR program package \cite{matthews2020cfour}, at the highly accurate CCSD(T)/ANO1 (tetra-atomics and smaller without Cl or S), CCSD(T)/cc-PVTZ (tetratomics and smaller with S or Cl), or Hartree-Fock/ANO0 (larger molecules) level of theory.

These Hamiltonians are stored in wavenumber units (cm$^{-1}$) commonly used by spectroscopists. We also include qubit encodings of dipole operators $\mu_{\{x,y,z\}}$, which are required for example in the calculation of transition probabilities \cite{Roggero19_linresp,jnane2021analog,ibe2022,sawaya2022notrap} when determining infrared spectra. Finally, in Appendix \ref{apx:reson} we report statistics for how closely-space the energy levels are in these Hamiltonians, as resonances (i.e. closely-space energy transitions) tend to be hard to simulate classically.

We include 3 encodings for each instance: unary (one-hot), standard binary, and Gray, as well as the pre-encoded ``bosonic'' Hamiltonians in mat2qubit~\cite{mat2qubit} string format. The following molecules are included.
\\

\begin{tabular}{ | p{0.25\linewidth} | p{0.65\linewidth}  |}
\hline
 Diatomics & BeH, BH, CH, CO, F$_2$, HF \\ 
 \hline
 Linear Triatomics & HNC, HNO, C$_2$O, C$_2$H\\  
  \hline
 Bent Triatomics & CH$_2$, COH, H$_2$O, H$_2$S, H$_3$$^+$, HCO, NH$_2$, NOH, O$_3$, SF$_2$, SO$_2$\\  
  \hline
 Tetratomics & BF$_3$, BH$_3$, BHF$_2$, CH$_3$, CH$_3$$^{+}$, ClCOH, FCCF, H$_2$CC (vinylidene), H$_2$CO, H$_2$O$_2$, HCCH\\  
   \hline
   Larger Molecules & Allene (C$_3$H$_4$), Cyclopropene (C$_3$H$_4$), Ethylene Oxide (C$_2$H$_4$O), Propargyl cyanide (HC$_3$H$_2$CN)\\
   \hline
\hline
 (Auxiliary) & Qubit-encoded dipole operators $\hat \mu_x$, $\hat \mu_y$, $\hat \mu_z$; \\
            & Cartesian coordinates of molecule and harmonic normal modes; \\
            & VPT2 Calculated transition energies \\
 \hline
\end{tabular}

\section{Benchmarking discussion}\label{sec:bench}

As noted previously, HamLib is not itself a benchmarking suite. Rather, in conjunction with a set of computational tasks, it may be used to define proper benchmarks. Such computational tasks may include ``full'' algorithms such as eigenvalue finding or quantum dynamics, but also narrower subroutines such as estimation of expectation values or compilation routines \cite{parekh2016benchmarking,chen2022veriqbench,li2022qasmbench,Lubinski21_bench,cornelissen2021scalable,tomesh2022supermarq,yeter2021benchmarking,mccaskey2019quantum,wu2023vscore}.

In this section, we discuss the various types of benchmarks that would benefit from a diverse library of Hamiltonians. At its heart, the purpose of benchmarking  is to make \textit{comparisons}---answering questions such as ``Which of these computers completes the task more quickly?" or ``Will this new algorithm yield more accurate results than previous state-of-the-art?" We distinguish between three categories. First, benchmarks that allow one to compare quantum algorithms (independent of hardware choice); second, those that compare different quantum hardware platforms including error-prone devices; and third, those that compare compilation and top-of-stack tasks. 
Though these three can be considered conceptually independent to some extent, many benchmarking tasks will fall under multiple categories.

Any benchmark involves the comparison of computational resources, accuracy, or both. In the former case, one studies the computational time (related to quantum circuit depth and circuit repetitions) and/or space requirements (number of qubits). Often, there are time-space trade-offs, implying that the best algorithmic choice is hardware-dependent. In the case of studying accuracy, one instead compares which computational approach leads to lower errors. Below, we discuss several possible tasks for which one might define benchmarks using HamLib. Many related quantum benchmarks have been previously proposed in the literature, as discussed in Section \ref{sec:attributes}.

Finally, we note that the above tasks may be used either for comparing quantum algorithms against each other (``quantum-vs-quantum''), or for comparing quantum algorithms against classical algorithms (``quantum-vs-classical''). Both types of comparisons are valuable and our intention is for HamLib to be used in both.

\subsection{Benchmarking quantum algorithms}

Here we discuss benchmarks designed to study quantum algorithms, independently of the quantum hardware used. The primary early use cases we envision for HamLib are eigenvalue finding and quantum dynamics. As mentioned above, these may be analyzed both in terms of solution quality (e.g. accuracy) and in terms of computational resources. 

Eigenvalue finding is a common task, both in quantum problems (for which ground or excited states are of interest) and classical optimization (where an extremal eigenvalue corresponds to the optimal value). For extremal eigenvalues (i.e. ground states), several quantum algorithms have been developed. Perhaps most well-studied for near-term hardware are the variational quantum eigensolver (VQE) \cite{mcclean2016theory} and the quantum approximate optimization algorithm (QAOA) \cite{farhi2014quantum}. Both of these involve a plethora of algorithmic choices. Other quantum algorithms for this task include adiabatic quantum optimization (or adiabatic state preparation) \cite{farhi2000adiabatic}, quantum imaginary-time evolution (QITE) \cite{motta2020determining}, feedback-based quantum algorithms \cite{magann2022}, 
and quantum subspace methods \cite{parrish2019qfd,Klymko2022}. 
Finding excited (non-extremal) eigenvalues requires either distinct quantum algorithms or modifications of ground state algorithms. Some notable approaches to excited state finding include the folded spectrum method \cite{wang1994folded,mcclean2016theory}, quantum variational deflation \cite{higgott2019vqd}, and quantum equations of motion \cite{ollitrault2020eom}; the unmodified versions of QITE and quantum subspace methods naturally provide estimates to excited states.

Notably, each algorithmic component may be benchmarked separately. For VQE and QAOA specifically, there are arguably three main components to consider. First, one may implement different quantum circuit ansatzae, also known as parameterized quantum circuits \cite{cao2019chemrev,Grimsley2019a,
xia2020qcc,zhang2021mutual,gard2020efficient,leone2022practical}. Second, one may compare different classical optimizers \cite{mcclean2016theory,guerreschi2017practical,singh2022optimizes}. Third, one may compare methods for calculating expectation values $\langle \psi|H|\psi \rangle$, which is often a very time-consuming subroutine 
\cite{Izmaylov2020,Gokhale2020,Crawford2021,Huang2020shadow,Klymko2022,Huggins2021}.

Simulating quantum dynamics---i.e. implementing the exponential $e^{-iH\Delta t}$ of a Hamiltonian---is a task distinct from eigenvalue finding and for which there are many algorithmic choices to be made. The most space-efficient approach is to use product formulas such as Suzuki-Trotter decompositions, for which one may consider different orders as well as more complex approaches such as randomized product formulas
\cite{lloyd1996universal,childs2019random,childs2021theory}. If one is primarily concerned with asymptotic scaling, as will often be the case when considering long-term hardware, there are several more advanced Hamiltonian simulation methods that require ancilla qubits \cite{low2017optimal,low2022trotter,miessen2023qdyn}. Notably, exponentiation is a key subroutine of other algorithms, including the above-mentioned quantum subspace methods and some quantum machine learning protocols.

For the algorithms discussed above, the choice of \textit{encoding} is very important and may be studied as a standalone consideration. Fermionic problems, bosonic and vibrational problems, and discrete combinatorial problems all involve rich choices of encoding \cite{cao2019chemrev,seeley2012bk,sawaya2020resource,sawaya2020connectivity,sawaya2022dqir, Miller2023}. Indeed, simply changing the encoding while leaving the rest of the quantum algorithm unchanged is a valuable study in itself.

We briefly touch on quantum machine learning (QML), which is conceptually distinct from eigenvalue finding and quantum dynamics. Some important machine learning tasks include classification, regression, clustering, and dimensionality reduction. Though such tasks were originally proposed in the context of classical data such as images or user data, some recent work has gone into the ``quantum quantum'' version of machine learning, whereby a quantum algorithm is used to analyze quantum data \cite{huang2021power}. In principle, some subsets of HamLib may be used in benchmarking for such tasks, as has already been proposed in QML for chemistry \cite{sajjan2022quantum}.

\subsection{Benchmarking quantum hardware}

In this section, we discuss procedures designed to benchmark specific quantum devices, especially near-term noisy devices \cite{preskill2018nisq}. %
Our hope is that HamLib spawns a set of benchmarks meant to complement the randomized benchmarking protocols that are often used in the industry \cite{knill2008randomized}. Such hardware-to-hardware comparisons are one major motivation behind benchmarking in classical computing, for example, when evaluating the performance of a new generation of processor or supercomputer (see Section \ref{sec:related}). 

The most obvious benchmarks in this category involve running eigenvalue finding and Hamiltonian simulation on noisy hardware, to determine which quantum device yields the most accurate result. Though protocols such as quantum volume \cite{cross2019validating,magesan2012characterizing,knill2008randomized,erhard2019cyclic} are valid, transferable, and objective metrics, it is unclear how such metrics relate to the accuracy of a practical workload. It could be that, when applied to scientifically relevant problems, some algorithms are less (or more) sensitive to noise than expected. For instance, it is not implausible that in some instances doubling the decoherence rate will lead to a relatively small reduction in accuracy for some problems. %
If there is such robustness to noise in a given device, then it is useful to understand its source and to know for which applications it occurs.  It is worth mentioning that studying energy consumption of the quantum device will be an important future direction as well \cite{feng2007green500,berger2021climate}.

Another class of benchmarks related to hardware involves the choice of encoding a logical qubit into physical qubits. This may involve a choice as simple as rotating each individual qubit basis, to mitigate e.g. amplitude damping noise. Or the consideration may be more complex, such as choosing between different decoherence-free subspaces \cite{lidar2003decoherence}. Again, it is plausible that the best choice of qubit encoding is problem-dependent, which could be studied using the HamLib library.

Finally, error mitigation protocols form an important area of study. There have been several proposed methods for mitigating errors on NISQ hardware \cite{endo2018errmitig,cao2021nisqerr,huggins2021virtualdistill}. Benchmarking such procedures for the different problem classes of HamLib may allow one to understand how and whether to apply such techniques in real applications.

\subsection{Benchmarking compilation and the full stack}

Here we discuss benchmarking of the quantum computing ``stack,'' especially compilation routines \cite{javadiabhari2014scaffcc,khalate2022llvm,kharkov2022arlinebench}. We place such efforts in a qualitatively separate category because in principle they do \textit{not} require real simulations nor real hardware. In other words, this section discusses benchmarks that do not involve interacting with the actual Hilbert space (whether via quantum hardware or via classical simulators).

Notably, though noisy hardware is likely to be the first testing environment for HamLib, we also intend for HamLib to be used with fully error-corrected hardware in mind. Among other differences, the gate set in error-corrected hardware tends to be different than that used in NISQ computers.
Especially important is resource estimation for error-corrected quantum computing. Useful studies of this type have been performed for chemistry, where estimates for required T gate counts have been determined for 
classically intractable molecules \cite{reiher2017pnas,li2019electronic,von2021catalysis}. In such cases, because one is not performing the actual quantum computation, one must use informed estimates for quantities such as how many operations will be required to prepare a trial state for QPE and how much overlap this trial state will have with the ground state. Such studies, performed using HamLib, may help researchers begin to understand how required resources scale with problem size, when solving various problems on error-corrected quantum computers. 

Mapping problems to qubit representations is often a time-consuming procedure. For instance, fermionic problems must implement the Jordan-Wigner (or related) mapping in order to preserve commutation properties, and the classical steps required may be substantial when one has (not uncommonly) $O(N^4)$ terms to consider. As mentioned above, analogous mappings are required for vibrational and bosonic problems, as well as classical problems. There is likely much room for efficiency improvement in generating qubit representations, for example via pre-processing, memoization, and parallelization.

Methods for quantum circuit construction and optimization may be benchmarked as well \cite{kliuchnikov2013optimization,abdessaied2014quantum,nam2018automated,wille2019qiskit,sivarajah2020t,pointing2021optimizing,xu2022quartz,paykin2023pcoast,schmitz2023optimization,cirq_developers_2022_7465577}. Both the speed of the compilation and its quality must be considered, where circuit depth might be one quality criterion. As an example, there are a plethora of choices for compiling circuits for Hamiltonian exponentiation, each of which has different accuracy, space, and depth trade-offs  \cite{duncan2020graph,schmitz2021graph,kharkov2022arlinebench,peham2022equivalence,khesin2023graphical}. 
Further, there are multiple ways to decompose a given unitary \cite{barenco1995elementary}, including via randomized compilation \cite{wallman2016randomizedcompilation}, each of which might be considered by the compiler. Gate fusions and cancellations may be handled with different approaches that may need to be compared. It will be interesting to see the extent to which classical compilation, a mature field of many decades, will bring techniques to bear on quantum circuit compilation.

Finally, we note that there are many aspects of running the quantum stack that are intimately related to a particular hardware design. When two-qubit gates are available only between physically adjacent qubits (which is the case for most hardware types), gate scheduling protocols \cite{metodi2006scheduling,guerreschi2018two} must be benchmarked and the choice of qubit topology (i.e. connectivity) intimately affects performance \cite{holmes2020connect}. As noted above, multiple choices for decompositions into native gates may be available as well, some of which may lead to more optimal circuits than others. And further into the future, certain aspects of the quantum-classical interface as it relates to quantum error correction ought to be optimized, including the analysis of syndrome measurements \cite{roffe2019qec}. 
As hardware-software codesign \cite{li2021co,wintersperger2022qpu} becomes more prevalent in quantum computer development, the HamLib dataset may play a role in guiding many hardware design choices.

\section{Concluding remarks}\label{sec:concl}

We have introduced a wide-ranging dataset of qubit Hamiltonians, for a variety of quantum problems including condensed matter models, electronic structure, and vibrational structure, as well as classical problems including Max-k-SAT, Max-k-Cut, and the traveling salesperson problem. We have left actual benchmarking studies to future work, using this first work to only introduce and characterize the dataset.

A primary purpose of this work was to provide the community with a set of problem instances that can be used directly and immediately, without the need to learn and implement the tedious multi-step processes required. Preparing electronic and vibrational structure instances for real-world molecules, for instance, involves significant domain knowledge and many software steps. Further, our inclusion of problems with real-world parameters and attributes (electronic structure, vibrational structure, TSP for European cities) is meant to encourage the community to continue its focus on problems instances that more closely relate to industrial impact.

HamLib may be used for several distinct types of tests: (a) simulating small problems on current noisy quantum computers, (b) classically emulating quantum computers, or (c) studying processes for which the quantum circuit itself does not need to be simulated, \textit{e.g.} for compilation procedures up to hundreds or thousands of qubits. These Hamiltonians may be implemented in conjunction with a range of algorithm classes for which new algorithmic approaches need to be tested or hyperparameters tuned: VQE, QAOA, adiabatic quantum state preparation, Krylov-subspace methods, and various long-term Hamiltonian simulation methods.

Quality classical datasets have long been an indispensable component of benchmarking for the design of supercomputers, processors, linear algebra methods, and AI algorithms. We hope that this dataset fills a similar role for the quantum computing community.

\section*{Acknowledgements}
We thank RoaRQ consortium for useful discussions on dataset coverage, Shrihan Agarwal for useful discussions on TSP, Timothy Proctor for manuscript feedback, and Neil Mehta for testing portions of the dataset. 
We acknowledge support from Sandia National Laboratories’ Laboratory Directed Research and Development Program under the Truman Fellowship. Sandia National Laboratories is a multimission laboratory managed and operated by National Technology \& Engineering Solutions of Sandia, LLC, a wholly owned subsidiary of Honeywell International Inc., for the U.S. Department of Energy’s National Nuclear Security Administration under contract DE-NA0003525. This paper describes objective technical results and analysis. Any subjective views or opinions that might be expressed in the paper do not necessarily represent the views of the U.S. Department of Energy or the United States Government. 
This material is based upon work supported by the U.S. Department of Energy (DOE), Office of Science, National Quantum Information Science Research Centers, Quantum Systems Accelerator.  Additional support is acknowledged from DOE, Office of Science, Office of Advanced Scientific Computing Research, Accelerated Research in Quantum Computing, Fundamental Algorithmic Research in Quantum Computing.
This research used resources of the National Energy Research Scientific Computing Center (NERSC), a U.S. Department of Energy Office of Science User Facility located at Lawrence Berkeley National Laboratory, operated under Contract No. DE-AC02-05CH11231. This included the use of NERSC award DDR-ERCAP0030342.  
The authors would like to acknowledge the use of the University of Oxford Advanced Research Computing (ARC) facility in carrying out this work (\url{http://dx.doi.org/10.5281/zenodo.22558}). 
NMT is grateful for support from NASA Ames Research Center.  We acknowledge funding from the NASA ARMD Transformational Tools and Technology (TTT) Project. DMD acknowledges financial support by the EPSRC Hub in Quantum Computing and Simulation
(EP/T001062/1). %
D.B.N. was supported by NASA Academic Mission Services, Contract No. NNA16BD14C.
D.P.T. acknowledges support from the Robert A. Welch Foundation, Grant no. A-2049-20230405. Portions of this research were conducted with high-performance research computing resources provided by Texas A\&M University HPRC.
\bibliographystyle{alpha}
\bibliography{refs}
\newpage

\appendix

\section{HamLib code snippets}\label{app:snip}

We provide a collection of Python functions that can be used to interact with the HamLib dataset. \textcolor{black}{The code snippets are also available via:
\begin{center}
\url{https://portal.nersc.gov/cfs/m888/dcamps/hamlib/hamlib_snippets.py}
\end{center}}

or

\begin{center}
\url{https://github.com/Azulene-Labs/hamlib_functions}.
\end{center}

The code %
depends on domain-specific Python packages, including NetworkX \cite{networkx}, OpenFermion \cite{openfermion}, and mat2qubit \cite{mat2qubit}, and some other widely used libraries. We specifically assume that the following imports have been made before using the functions in Sections~\ref{app:h5struct} and \ref{app:h5data}:

\begin{lstlisting}[language=Python]
import networkx as nx
import mat2qubit as m2q
import openfermion as of
import h5py
import numpy as np
import re
from qiskit.quantum_info import SparsePauliOp
\end{lstlisting}

\subsection{Loading HDF5 file structure}\label{app:h5struct}

We recommend using the function \texttt{print\_hdf5\_structure} to inspect the path tree structure of the datasets stored in the HamLib HDF5 files and \texttt{get\_hdf5\_keys} to return a list of keys to all datasets stored in a HamLib HDF5 file. 
This list of keys can be used to extract data from the HDF5 file using the functions provided in Section~\ref{app:h5data}.

Both \texttt{print\_hdf5\_structure} and \texttt{get\_hdf5\_keys} make use of the decorator function\linebreak \texttt{parse\_through\_hdf5}.

\begin{lstlisting}[language=Python]
def parse_through_hdf5(func):
    """Decorator function that iterates through an HDF5 file and performs
    the action specified by `func` on the internal and leaf nodes in the HDF5 file."""
    def wrapper(obj, path='/', key=None):
        if type(obj) in [h5py._hl.group.Group, h5py._hl.files.File]:
            for ky in obj.keys():
                func(obj, path, key=ky, leaf=False)
                wrapper(obj=obj[ky], path=path + ky + '/', key=ky)
        elif type(obj) == h5py._hl.dataset.Dataset:
            func(obj, path, key=None, leaf=True)
    return wrapper

\end{lstlisting}

\begin{lstlisting}[language=Python]
def print_hdf5_structure(fname_hdf5: str):
    """Print the path structure of the HDF5 file.

    Args
    ----
        fname_hdf5 (str): full path where HDF5 file is stored
    """

    @parse_through_hdf5
    def action(obj, path='/', key=None, leaf=False):
        if key is not None:
            print((path.count('/')-1)*'\t', '-', key, ':', path + key + '/')
        if leaf:
            print((path.count('/')-1)*'\t', '[^^DATASET^^]')

    with h5py.File(fname_hdf5, 'r') as f:
        action(f['/'])
\end{lstlisting}

\begin{lstlisting}[language=Python]
def get_hdf5_keys(fname_hdf5: str):
    """Get a list of keys to all datasets stored in the HDF5 file. 
    
    Args
    ----
        fname_hdf5 (str): full path where HDF5 file is stored
    """
    all_keys = []

    @parse_through_hdf5
    def action(obj, path='/', key=None, leaf=False):
        if leaf is True:
            all_keys.append(path[:-1])

    with h5py.File(fname_hdf5, 'r') as f:
        action(f['/'])

    return all_keys
\end{lstlisting}

\subsection{Loading HDF5 data}\label{app:h5data}
The five functions listed below can be used to load data stored at a specific key in the HDF5 file back into an appropriate Python objects, which respectively are a NetworkX graph, a dictionary of grid positions, OpenFermion operators (or qiskit SparsePauliOp operators), mat2qubit operators and a list of clauses.

\begin{lstlisting}[language=Python]
def read_graph_hdf5(fname_hdf5: str, key: str):
    """Read networkx graphs from HDF5 file at specified key. Returns a single networkx graph.
    """
    with h5py.File(fname_hdf5, 'r') as f:
        G = nx.Graph(list(np.array(f[key])))

    return G
\end{lstlisting}

\begin{lstlisting}[language=Python]
def read_gridpositions_hdf5(fname_hdf5: str, key: str):
    """Read grid positions, stored as attribute of each networkx graph from HDF5 file at specified key. Returns grid positions of nodes associated with a single graph.
    """
    with h5py.File(fname_hdf5, 'r') as f:
        dataset = f[key]
        gridpositions_dict = dict(dataset.attrs.items()) 

    return gridpositions_dict
\end{lstlisting}

\begin{lstlisting}[language=Python]
def read_openfermion_hdf5(fname_hdf5: str, key: str, optype=of.QubitOperator):
    """Read any openfermion operator object from HDF5 file at specified key.
    'optype' is the op class, can be of.QubitOperator or of.FermionOperator.
    """
    with h5py.File(fname_hdf5, 'r', libver='latest') as f:
        op = optype(f[key][()].decode("utf-8"))

    return op
\end{lstlisting}

\begin{lstlisting}[language=Python]
def read_qiskit_hdf5(fname_hdf5: str, key: str):
    """
    Read the operator object from HDF5 at specified key to qiskit SparsePauliOp
    format.
    """
    def _generate_string(term):
        # change X0 Z3 to XIIZ
        indices = [
            (m.group(1), int(m.group(2))) 
            for m in re.finditer(r'([A-Z])(\d+)', term)
        ]
        return ''.join(
            [next((char for char, idx in indices if idx == i), 'I')
             for i in range(max(idx for _, idx in indices) + 1)]
        )

    def _append_ids(pstrings):
        # append Ids to strings
        return [p + 'I' * (max(map(len, pstrings)) - len(p)) for p in pstrings]

    with h5py.File(fname_hdf5, 'r', libver='latest') as f:
        pattern = r'([\d.]+) \[([^\]]+)\]'
        matches = re.findall(pattern, f[key][()].decode("utf-8"))

        labels = [_generate_string(m[1]) for m in matches]
        coeffs = [float(match[0]) for match in matches]
        op = SparsePauliOp(_append_ids(labels), coeffs)
    return op
\end{lstlisting}

\begin{lstlisting}[language=Python]
def read_mat2qubit_hdf5(fname_hdf5: str, key: str):
    """Returns mat2qubit's qSymbOp operator from HDF5 file at specified key."""
    with h5py.File(fname_hdf5, 'r') as f:
        op = m2q.qSymbOp(f[key][()].decode("utf-8"))

    return op
\end{lstlisting}

\begin{lstlisting}[language=Python]
def read_clause_list_hdf5(fname_hdf5: str, key: str):
    """Read clause list from HDF5 file at specified key.Returns clause list in DIMACS format."""
    clause_list = []
    with h5py.File(fname_hdf5, 'r') as f:
        for clause in list(np.array(f[key])):
            clause_list.append([v for v in clause])

    return clause_list
\end{lstlisting}

\subsection{Loading and unzipping to memory}

For certain workflows, especially ones targeting some of the smaller subdatasets of HamLib, we expect that it might be useful to load the data directly from the webportal into a Python environment. This has an additional benefit of code portability.

Alternatively one may download the zip files \href{https://portal.nersc.gov/cfs/m888/dcamps/hamlib/}{directly}. We recommend manually downloading the larger ($>$100 MB) files of HamLib, as the uncompressed data is often 10x the size of the compressed data. Additionally, we note that for some users a manual download may be the only option---during testing, some users were unable to access HamLib via the commands below, due to virtual private network (VPN) or related issues.

The code snippet below provides an example of how this can be achieved.
We note that this approach loads the full zip file into memory and thus
might be slow for some of the larger datasets.

\begin{lstlisting}[language=Python]
import zipfile
import requests
from io import BytesIO

url = 'https://portal.nersc.gov/cfs/m888/dcamps/hamlib/test/graph/test_graph.zip'

r = requests.get(url, stream=True)
z = zipfile.ZipFile(BytesIO(r.content))

hdf5_filename = z.namelist()[0]
G = read_graph_hdf5(z.open(hdf5_filename, 'r'), "test_graph")
\end{lstlisting}

\subsection{Operator manipulation}

The two functions below are used for reducing the size of a Hamiltonian, either by removing indices or removing terms. The first function will remove indices; perhaps its most important use is removing higher-energy orbitals in HamLib's electronic structure problems. The second function will remove smaller terms from an operator.

\begin{lstlisting}[language=Python]
def remove_qindices(op, inds_to_remove):
    """Removes indices (i.e. removes qubits or fermionic modes) from operator. Any term
    containing the given indices are removed.

    Args:
        op (inhereting from SymbolicOperator): Operator with indices to be removed

    Returns:
        new_op (inhereting from SymbolicOperator): New operator with indices removed
    """

    assert isinstance(op, SymbolicOperator)
    # assert inds_to_remove is iterable
    assert hasattr(inds_to_remove, '__iter__'), "Input is not iterable"

    new_op = copy.deepcopy(op)

    for term in op.terms:
        for factor in term:
            if factor[0] in inds_to_remove:
                new_op.terms.pop(term)
                break

    return new_op

\end{lstlisting}

\begin{lstlisting}[language=Python]
def remove_smaller_values(op,thresh):
    """Removes absolute values below given threshold"""

    assert isinstance(op, SymbolicOperator)

    new_op = copy.deepcopy( op )

    for term in op.terms:
        if abs(op.terms[term]) < thresh:
            new_op.terms.pop(term)

    return new_op

\end{lstlisting}

\section{Qubit Hamiltonian metadata distributions for subdatasets}\label{app:figures}
\begin{figure}[h]
    \centering
    \includegraphics[width=\linewidth]{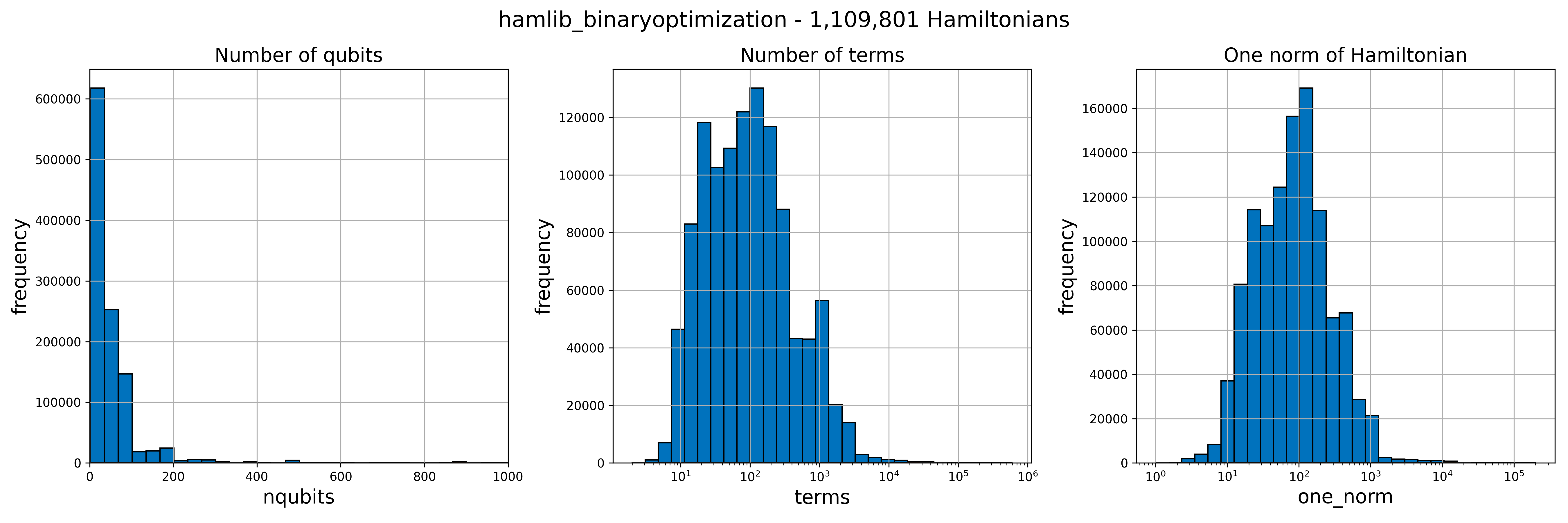}
    \caption{\textcolor{black}{Distribution of the number of qubits (\textbf{left}), number of terms (\textbf{middle}), and one norms (\textbf{right}) over all qubit Hamiltonians in the binary optimization subdataset of HamLib.}}
    \label{fig:hamlib-binaryopt-distribution}
\end{figure}

\begin{figure}[h]
    \centering
    \includegraphics[width=\linewidth]{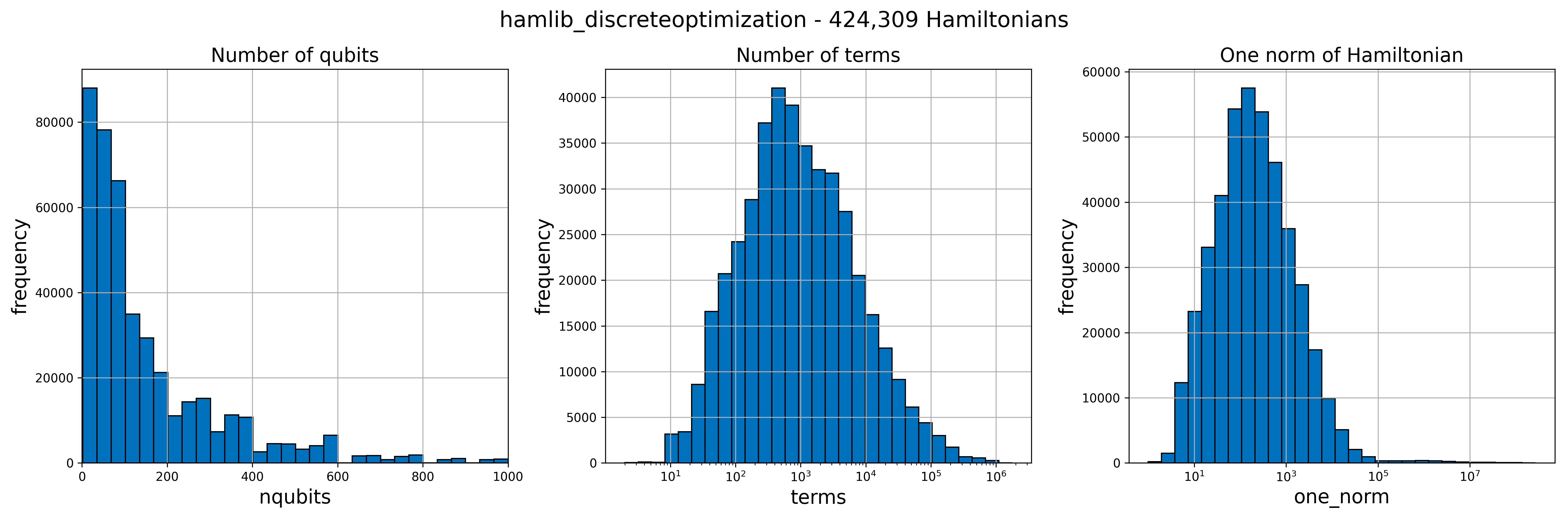}
    \caption{\textcolor{black}{Distribution of the number of qubits (\textbf{left}), number of terms (\textbf{middle}), and one norms (\textbf{right}) over all qubit Hamiltonians in the discrete optimization subdataset of HamLib.}}
    \label{fig:hamlib-discreteopt-distribution}
\end{figure}

\begin{figure}[h]
    \centering
    \includegraphics[width=\linewidth]{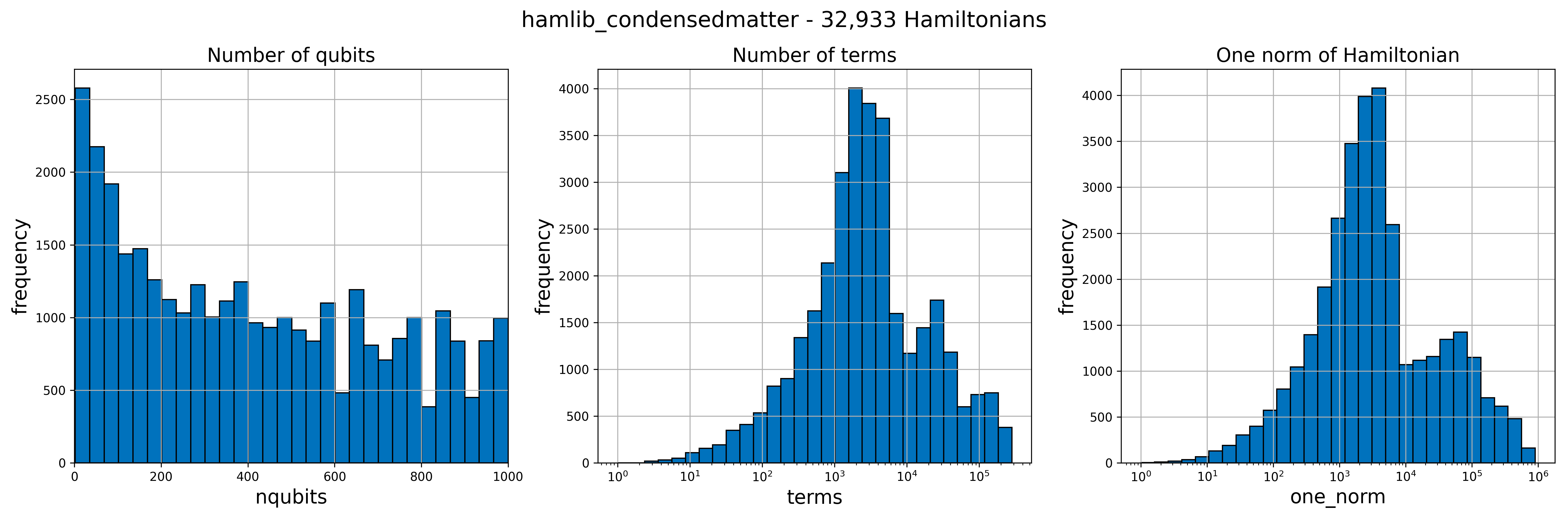}
    \caption{\textcolor{black}{Distribution of the number of qubits (\textbf{left}), number of terms (\textbf{middle}), and one norms (\textbf{right}) over all qubit Hamiltonians in the condensed matter subdataset of HamLib.}}
    \label{fig:hamlib-condensedmatter-distribution}
\end{figure}

\begin{figure}[h]
    \centering
    \includegraphics[width=\linewidth]{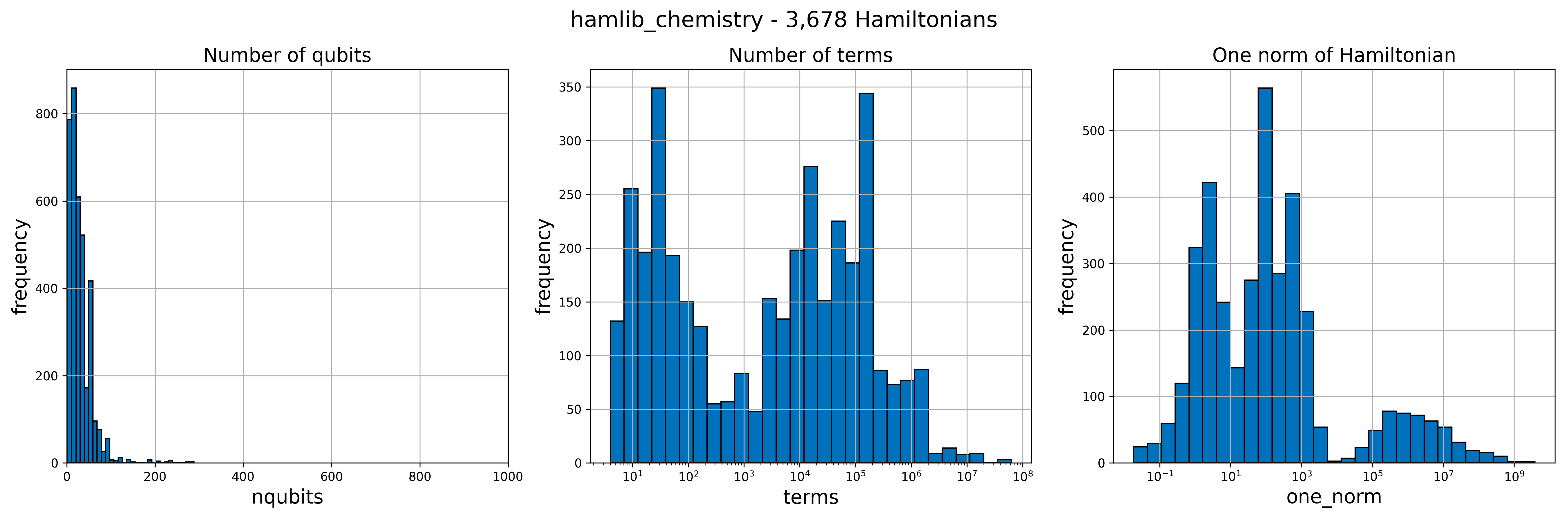}
    \caption{\textcolor{black}{Distribution of the number of qubits (\textbf{left}), number of terms (\textbf{middle}), and one norms (\textbf{right}) over all qubit Hamiltonians in the chemistry subdataset of HamLib.}}
    \label{fig:hamlib-chemistry-distribution}
\end{figure}

Figures~\ref{fig:hamlib-binaryopt-distribution}--\ref{fig:hamlib-chemistry-distribution} show an overview of the metadata (`\texttt{nqubits}`, `\texttt{terms}`, and `\texttt{one$\_$norm}') for respectively the binary optimization, discrete optimization, condensed matter and chemistry subdatasets of HamLib. These give a sense for the distribution of Hamiltonian properties present in these datasets.

\section{Number of frozen electrons}\label{sec:frozen_elec}

For our electronic structure Hamiltonians, the frozen orbitals counts are given in Table \ref{tbl:frozen_orbitals}. The number of frozen electrons is twice this value. These values are required to determine how many effective electrons to include in a simulation. For example, the physical molecule BH contains 6 electrons. Because the table shows that 2 electrons were frozen to produce the Hamiltonian, the user should implement HamLib's BH Hamiltonian with 6-2=4 electrons.

\begin{table}[h]
\centering
\begin{tabular}{l c c}
\hline
\textbf{Molecule} & \textbf{No. Frozen Orbitals} & \textbf{No. Frozen Electrons} \\
\hline
B\textsubscript{2}             & 2 & 4 \\
BH                            & 1 & 2 \\
Be\textsubscript{2}            & 2 & 4 \\
BeH                           & 1 & 2 \\
C\textsubscript{2}             & 2 & 4 \\
CH                            & 1 & 2 \\
CoH                           & 1 & 2 \\
Cr\textsubscript{2}            & 2 & 4 \\
CrO                           & 1 & 2 \\
CuC                           & 1 & 2 \\
F\textsubscript{2}             & 2 & 4 \\
FeC                           & 1 & 2 \\
H\textsubscript{2}             & 0 & 0 \\
HF                            & 1 & 2 \\
Li\textsubscript{2}            & 2 & 4 \\
LiH                           & 1 & 2 \\
MnN                           & 1 & 2 \\
N\textsubscript{2}             & 2 & 4 \\
NH                            & 1 & 2 \\
Na\textsubscript{2}            & 2 & 4 \\
NaLi                          & 1 & 2 \\
NiO                           & 1 & 2 \\
O\textsubscript{2}             & 2 & 4 \\
O\textsubscript{3}             & 3 & 6 \\
OH                            & 1 & 2 \\
ScC                           & 1 & 2 \\
ScO                           & 1 & 2 \\
TiH                           & 1 & 2 \\
VH                            & 1 & 2 \\
\hline
\end{tabular}
\caption{The number of frozen orbitals and frozen electrons in HamLib's electronic structure Hamiltonians.}
\label{tbl:frozen_orbitals}
\end{table}

\section{Resonances in vibrational Hamiltonians}\label{apx:reson}

For molecular vibrational Hamiltonians, Table \ref{tbl:resonances} shows how many pairs of transitions are within the specified energy threshold. Because near-degenerate transitions (\textit{i.e.} resonances) are difficult to treat with classical algorithms, the number of such transitions may be used as a rough proxy for classical hardness.
\\

\begin{table}[h]
\centering
\begin{tabular}{lrrrrr}
\\
            Molecule &  25 cm$^{-1}$ &  10 cm$^{-1}$ &  5 cm$^{-1}$ &  2 cm$^{-1}$ &  1 cm$^{-1}$  \\
\hline
      BH$_3$ &         148 &                  74 &        40 &        40 &        39 \\
   H$_2$CO &                   27 &                  9 &         7 &         3 &         2 \\
       BHF$_2$ &                        20 &              3 &         0 &         0 &         0 \\
 H$_2$CC &                       11 &         2 &         1 &         1 &         1 \\
   H$_2$O$_2$ &                    31 &         21 &        10 &         1 &         0 \\
          BF$_3$ &                    106 &         68 &        44 &        44 &        44 \\
         CH$_3$ &                        88 &         58 &        38 &        34 &        34 \\
 CH$_3^{+}$ &                   117 &         57 &        45 &        36 &        34 \\
    ClCOH &                       20 &          8 &         3 &         1 &         1 \\
    FCCF &                    460 &        252 &       184 &       102 &        87 \\
     HCCH &                     150 &        111 &        72 &        56 &        48 \\
      C$_2$H &                     25 &         21 &        21 &        17 &        11 \\
      C$_2$O &                             27 &         19 &        13 &        13 &        11 \\
     CH$_2$ &                          0 &          0 &         0 &         0 &         0 \\
       COH &                           1 &          0 &         0 &         0 &         0 \\
       H$_2$O &                3 &          3 &         2 &         0 &         0 \\
       H$_2$S &                 4 &          1 &         0 &         0 &         0 \\
      H$_3^{+}$ &                          7 &          7 &         7 &         7 &         7 \\
      HCO &                  1 &          0 &         0 &         0 &         0 \\
          HNC &                    21 &         17 &        11 &        11 &        11 \\
             HNO &                        1 &          1 &         0 &         0 &         0 \\
                 NH$_2$ &                      2 &          0 &         0 &         0 &         0 \\
                   O$_3$ &                      0 &          0 &         0 &         0 &         0 \\
               SF$_2$ &                    9 &          0 &         0 &         0 &         0 \\
                SO$_2$ &                    0 &          0 &         0 &         0 &         0 \\
                  NOH &          0 &          0 &         0 &         0 &         0 \\
             BeH &                    0 &          0 &         0 &         0 &         0 \\
                BH &                      0 &          0 &         0 &         0 &         0 \\
               CH &               0 &          0 &         0 &         0 &         0 \\
                 CO &                     0 &          0 &         0 &         0 &         0 \\
                  F$_2$ &                   0 &          0 &         0 &         0 &         0 \\
                  HF &                     0 &          0 &         0 &         0 &         0 \\
            Allene &          3243 &        1707 &      1224 &       951 &       791 \\
       Cyclopropene &               2827 &                 1030 &       512 &       181 &        90 \\
 Ethylene oxide &            4090 &       1902 &       883 &       349 &       174 \\
          Propargyl Cyanide &             6611 &       2760 &      1482 &       564 &       294 \\
\hline
\end{tabular}
\caption{Number of resonances in HamLib's vibrational structure Hamiltonians.}
\label{tbl:resonances}
\end{table}

\end{document}